\documentclass[12pt]{article}
\usepackage{a4}
\usepackage{epsfig}
\newcommand{\be}{\begin{equation}}
\newcommand{\ee}{\end{equation}}
\newcommand{\ba}{\begin{eqnarray}}
\newcommand{\ea}{\end{eqnarray}}
\newcommand{\MS}{\ensuremath{\overline{\mbox{MS}}}}
\newcommand{\psix}{\ensuremath{{\cal O}(p^6)}}
\newcommand{\pfour}{\ensuremath{{\cal O}(p^4)}}

\begin{document}
\begin{titlepage}
\begin{flushright}
LU TP 99-15\\
hep-ph/9907264\\
July 1999
\end{flushright}
\vfill
\begin{center}
{\Large\bf
Two-point Functions at Two Loops in Three Flavour Chiral Perturbation
Theory$^\dagger$}\\[2cm]
{\bf Gabriel Amor\'os$^{a,b}$, Johan Bijnens$^a$ and Pere Talavera$^a$}\\[1cm]
{$^a$ Department of Theoretical Physics 2, Lund University,\\
S\"olvegatan 14A, S22362 Lund, Sweden}\\[1cm]
{$^b$ Department of Physics, P.O. Box 9\\
FIN-00014 University of Helsinki, Finland}
\end{center}
\vfill
\begin{abstract}
The vector and axial-vector two-point functions are calculated to
next-to-next-to-leading order in Chiral Perturbation Theory for
three light flavours. We also obtain expressions at the same
order for the masses, $m_\pi^2$, $m_K^2$ and $m_\eta^2$, and the decay
constants, $F_\pi$, $F_K$ and $F_\eta$. We present some numerical
results after a simple resonance estimate of some of the new ${\cal O}(p^6)$ constants.
\end{abstract}
\vfill
\vfill
\footnoterule
{$^\dagger$\small Work
supported in part by TMR, EC-Contract No. ERBFMRX-CT980169
(EURO\-DA$\Phi$NE).}
\end{titlepage}

\section{Introduction}

With the new collider facilities, upcoming experiments 
will bring higher statistics data samples into the low energy regime.
Due to their accuracy 
higher order calculations are needed
to update the theoretical prediction for the measurements. In this frame 
the calculation of the two-point functions at next-to-next-to-leading 
order (NNLO) at low energies has become necessary.
These provide us with the 
pseudoscalar masses and the
decay constants, which are needed input for most other quantities.
The two-point Green functions are also basic tools in the study 
of the strong
interaction. They form the basis for a series of very useful sum rules
starting with the Weinberg \cite{WSR} and DMO \cite{DMO} sum rules 
(we refer to \cite{booknarison} and references therein for a 
more complete discussion about sum rules).

In this work we are concerned with the low energy regime of QCD.
We will study the two-point functions with Chiral Perturbation 
Theory (CHPT), valid for energies 
below the first resonance ($E \sim m_\rho$) and describing the strong
interactions using the pseudoscalar octet as the basic fields. This
is by now 
a fairly developed field. We refer to \cite{badhonnef} for 
reviews and various abstracts on recent works.

For a future study of various sum rules, we present the vector 
and axial-vector two-point functions at NNLO in three flavour 
CHPT, in the limit of unbroken 
isospin. Four of the six basic correlators have been calculated
earlier \cite{KG1,KG4} and we have fully confirmed their results in the
vector-vector two point function and partially\footnote{
\label{footnote1} The authors of Refs.
\cite{KG1,KG4} use a different method to perform the sunset integrals
making comparison of those parts difficult. We agree on all the parts we could
check without converting their sunset functions to ours.}
in the axial-vector case.
The other two are new and complete the three flavour basis.
As a byproduct we also give the masses and decay constants to NNLO.

The interest in the NNLO calculation is beyond the 
precise measurement of the couplings and masses of the effective theory, 
it allows to test the convergence of the theory and provides
a more stringent check on the principle of resonance saturation
of the constants in the low energy chiral Lagrangian.
While this principle worked well at \pfour\ \cite{EG}
only a few tests at \psix\ have been done.
In this paper we estimate some of the constants 
appearing to two loops and check their effect on the full
\psix\ results including the loop contributions. We use 
the \MS-subtraction scheme and the recent 
classification for the \psix\ Lagrangian \cite{p6}.

Some applications to chiral sum rules for the isospin and hypercharge 
cases can be found in \cite{KG2,KG3}. We intend to return to that subject 
in a future publication.

The paper is organized as follows. In Sect. \ref{deftwo} the two-point
Green functions  are defined, followed by a short overview of CHPT 
in Sect. \ref{chpt}. The vector two-point function
is relatively easy since it only involves products of one-loop
integrals. Its calculation is described and results given in
Sect. \ref{vector}.
The masses and decay constants can be calculated in two ways. The masses
can be obtained
from the zero of the inverse pseudoscalar propagator or from the pole
in the axial-vector
two-point function. The decay constants can be directly 
determined from the residue of the poles in the axial-vector
two-point function or through the definition with 
the axial-vector-pseudoscalar function. 
We have checked that both methods agree
and the first is described in Sect. \ref{masses}. Finally the axial-vector
two-point functions are 
presented in Sect. \ref{axial}.
The new \psix\ constants appearing are estimated in Sect. \ref{estimates}
on the basis of Resonance Dominance.
In Sect. \ref{numerics} some results are presented, postponing
a more detailed analysis \cite{sumrules}.
And finally 
in Sect. \ref{conclusions} we discuss our main results.

We refer the lengthiest expressions and the more technical discussion of the
loop integrals and renormalization to appendices.
In App. \ref{appmass} and App. \ref{appdecay} we give the full
expressions for the masses and the decay constants. In App. \ref{axialvector}
we display the the axial-vector two point function components.
The loop integrals
are collected in App. \ref{loopint}.

\section{Definition of Two-point Functions}
\label{deftwo}

We calculate in CHPT the two-point functions of vector and axial-vector
currents. The quark currents are defined by
\be
V_\mu^{ij}(x) = \overline{q}^i \gamma_\mu q^j
\quad\mbox{and}\quad
A_\mu^{ij}(x) = \overline{q}^i \gamma_\mu\gamma_5 q^j\;,
\ee
where the indices
$i$ and $j$ run  over the  three light quark flavours,
$u$, $d$ and $s$.
Working in the isospin limit all SU(3) currents can be constructed
using isospin relations from
\ba
V_\mu^\pi(x) &=& \frac{1}{\sqrt{2}}\left(V_\mu^{11}(x)-V_\mu^{22}(x)\right)\,,
\nonumber\\
V_\mu^\eta(x) &=& \frac{1}{\sqrt{6}}\left(V_\mu^{11}(x)+V_\mu^{22}(x)
               -2V_\mu^{33}(x)\right)\,,
\nonumber\\
V_\mu^K(x) &=& V_\mu^{31}(x)\;.
\ea
We refer to these as isospin, hypercharge and kaon respectively.
E.g. the electromagnetic current corresponds to
\be
V_\mu^{{em}} =
\frac{e}{\sqrt{2}}V_\mu^\pi(x)+\frac{e}{\sqrt{6}}V_\mu^\eta(x)\;.
\ee

We calculate the two-point functions defined as
\be
\label{deftwop}
\Pi_{\mu\nu}^{Va}(q) \equiv
i\int d^4x\; e^{iq\cdot x}\;\langle 0|T(V_\mu^a(x)V_\nu^a(0))^\dagger|0\rangle \,,
\ee
for $a=\pi,\eta,K$. All other vector two-point functions can be constructed
from these using isospin relations.
Lorentz-invariance allows to express them
in a transverse, $\Pi^{(1)}$, and a longitudinal, $\Pi^{(0)}$, part
\be
\Pi_{\mu\nu}^{Va} = (q_\mu q_\nu-q^2 g_{\mu\nu})\Pi_{Va}^{(1)}(q^2)
 +q_\mu q_\nu \Pi_{Va}^{(0)}(q^2)\;.
\ee

Similar definitions and comments apply for the axial-vector currents.

The currents obey Ward identities and other symmetry relations;
for $m_u=m_d$ the vector currents are conserved
\be
\Pi_{V\pi}^{(0)}(q^2)=\Pi_{V\eta}^{(0)}(q^2)=0\;.
\ee
In the $SU(3)_V$ limit, the three vector two-point functions
reduce to the same expression. The same holds for the axial-vector currents. 
In the addition the last ones are conserved only when the relevant
quark masses vanish.

\section{Chiral Perturbation Theory}
\label{chpt}

Effective theories are a general tool in understanding a wide range
of physical processes, from high energy physics to superconductivity. 
In that frame
Chiral Perturbation Theory is a successful theory describing the strong
interaction at low energy. It is based on the existence of a mass-gap
in the hadronic spectrum, at low energies only the 
low mass states can be excited. Those are the Goldstone boson
particles and 
are the only states that are actually predicted
from first principles in QCD \cite{witten}. 
For the present status of the field we refer
to the listing of review articles and the abstracts in \cite{badhonnef} and to
some recent lectures \cite{CHPTlectures}.

For constructing the effective
action, the high energy states of the theory should be integrated out,
thus the Lagrangian describing 
processes at low energy consists of a series of operators involving
only Goldstone boson particles. 
These general operators should share the same symmetries as the 
\emph{basic} underlying theory \cite{weinberg,heiri},
in particular Lorentz invariance, local chiral symmetry,
parity and charge conjugation. 
Thereby in the following we use that the generating functional 
of both theories, QCD and CHPT, are the same
at low energies \cite{heiri}
\be
\exp\{iZ\} = \int {\cal D}q {\cal D}\bar{q} {\cal D}G_\mu 
\,\exp\bigg\{i\int d^4x {\cal L}_{QCD} 
\bigg\} = \int {\cal D}U \,\exp\bigg\{i\int d^4x {\cal L}_{eff}\bigg\}\,.
\ee
Following
this philosophy, the QCD effective Lagrangian
is given by a non-linear realization of chiral symmetry 
(see \cite{GL} and references therein). The lowest order in 
an expansion by quark masses and external momenta is 
\be
\label{lowest}
{\cal L}_2 = \frac{F_0^2}{4} \langle D_\mu U D^\mu U^\dagger 
+ U^\dagger \chi + \chi^\dagger U \rangle,
\ee
where 
\be
U(\phi) = u(\phi)^2 = \exp(i \sqrt{2} \Phi/F_0)\,,
\ee
parametrizes the pseudo-Goldstone bosons and
\be
\chi = 2 B_0 (s+ip)
\ee
is given in terms of the scalar and pseudoscalar sources $s$ and $p$. 
$\langle X \rangle = \mbox{tr}_{flavour}(X)$ and $U, s ,p$ are matrices in
flavour space.
Both $B_0$ and $F_0$ are constants not restricted by symmetry.
They are related
with the quark condensate and the meson decay constant respectively.
To respect local invariance the external sources are incorporated through
the covariant derivatives
\be
D_\mu U = \partial_\mu U -i r_\mu U + i U l_\mu \,,
\ee
and the field strength tensors
\be
\label{fieldstrength}
F_L^{\mu \nu} = \partial^\mu l^\nu -\partial^\nu l^\mu -i [l^\mu,l^\nu], \quad
F_R^{\mu \nu} = \partial^\mu r^\nu -\partial^\nu r^\mu -i [r^\mu,r^\nu]\,.
\ee

As mentioned above the purpose of this paper is to compute two-point Green
functions, 
the most straightforward way is to incorporate classical sources
in the effective action.
This reduces the calculation of any $n$-point Green function to the
evaluation of functional derivatives acting on the generating functional
\be
G^{(n)}(x_1,\ldots,x_n) = 
\frac{\delta^n}{\delta j(x_1) \ldots \delta j(x_n)} Z[J] 
\bigg\vert_{J=0}\,.
\ee
This allows for instance, to relate the chiral condensate with the
constant $B_0$ by taking
the derivative respect to the scalar sources.
This formalism is not only suitable for an easy calculation but also
allows to incorporate the electromagnetic,
weak interactions and the symmetry breaking through
the quark masses via the following identifications
\be
r_\mu = v_\mu + a_\mu, \quad
l_\mu=v_\mu - a_\mu, \quad
s= {\cal M} + \ldots \,,
\ee
where ${\cal M}$ stands for the diagonal quark mass matrix, 
$
{\cal M} = diag\,(m_u,m_d,m_s)\,. 
$

To get the desired chiral order ---\psix --- in our calculation we will deal 
with three kind of contributions:
$i)$ tree, one and two-loop diagrams involving vertices
of ${\cal L}_2$, $ii)$ tree and 1-loop graphs with vertices from
${\cal L}_2$ and from ${\cal L}_4$ given by
\ba
\label{l4}
{\cal L}_4 &=& L_1 \langle \nabla_\mu U^\dagger \nabla^\mu U \rangle^2 + 
L_2 \langle \nabla_\mu U^\dagger \nabla_\nu U \rangle 
 \langle \nabla^\mu U^\dagger \nabla^\nu U \rangle
\nonumber \\ \nonumber &&
+ L_3 \langle \nabla_\mu U^\dagger \nabla^\mu 
  U \nabla_\nu U^\dagger \nabla^\nu U \rangle
+ L_4 \langle \nabla_\mu U^\dagger \nabla^\mu U \rangle  \langle \chi^\dagger U + \chi U^\dagger \rangle
\\ \nonumber &&
+ L_5 \langle \nabla_\mu U^\dagger \nabla^\mu U 
  (\chi^\dagger U + \chi U^\dagger) \rangle
+ L_6 \langle \chi^\dagger U + \chi U^\dagger \rangle^2
\\ \nonumber &&
+ L_7 \langle \chi^\dagger U - \chi U^\dagger \rangle^2
+ L_8 \langle \chi^\dagger U \chi^\dagger U 
  +  \chi U^\dagger  \chi U^\dagger\rangle
\\ \nonumber &&
-i L_9 \langle F_R^{\mu \nu} \nabla_\mu U \nabla_\nu U^\dagger 
+ F_L^{\mu \nu} \nabla_\mu U^\dagger \nabla_\nu U \rangle
+ L_{10} \langle U^\dagger  F_R^{\mu \nu} U  F_{L \mu \nu} \rangle
\\  &&
+ H_1  \langle F_R^{\mu \nu}  F_{R \mu \nu} 
   + F_L^{\mu \nu}  F_{L \mu \nu} \rangle
+ H_2 \langle  \chi^\dagger  \chi \rangle\, 
\ea
and finally $iii)$ the tree graphs of ${\cal L}_6$. The latter was first
classified in \cite{fearing} and
recently in \cite{p6} a more restrictive general set was found.
We borrow in the following our notation from the last reference.
Notice that the terms $H_1$ and $H_2$ in Eq. (\ref{l4}) have no direct physical
meaning, their value depends on the precise way in which
the currents are defined in QCD. 
But once a consistent definition of a QCD current has been given they are 
defined unambiguously.

For later use we define the following quantities
\begin{eqnarray}
v^{\mu\nu} &=& \left(F_R^{\mu\nu} + F_L^{\mu\nu}\right)/2\,,
\nonumber\\ 
a^{\mu\nu} &=& \left(F_R^{\mu\nu} - F_L^{\mu\nu}\right)/2\,,
\nonumber\\
u_\mu &=& i\{u^\dagger(\partial_\mu-i r_\mu)u
  -u(\partial_\mu-i l_\mu)u^\dagger\}\,,
\nonumber\\
\Gamma_\mu &=& \frac{1}{2}\{u^\dagger(\partial_\mu-i r_\mu)u
  +u(\partial_\mu-i l_\mu)u^\dagger\}\,,
\nonumber\\
\chi_+&=&u^\dagger \chi u^\dagger+u \chi^\dagger u\,,
\nonumber\\
\nabla_\mu X &=& \partial_\mu X + \Gamma_\mu X - X \Gamma_\mu\,.
\end{eqnarray}

\section{The Vector Two-Point Functions}
\label{vector}

Within the framework of previous sections we can start to calculate the 
vector-vector two-point functions. The first contribution appears at 
$\pfour$. It has been calculated for the isospin case
in \cite{GL1} in SU(2) CHPT. The extension to $\psix$ in SU(3) CHPT 
has been done in \cite{KG1} for the isospin and the hypercharge case.
We have reproduced their results and in addition we present the kaon
vector two-point function as well here.

The calculation to $\psix$ presents no new difficulties 
besides being rather tedious
since only products of one-loop integrals appear.

\begin{figure}
\begin{center}
\epsfig{file=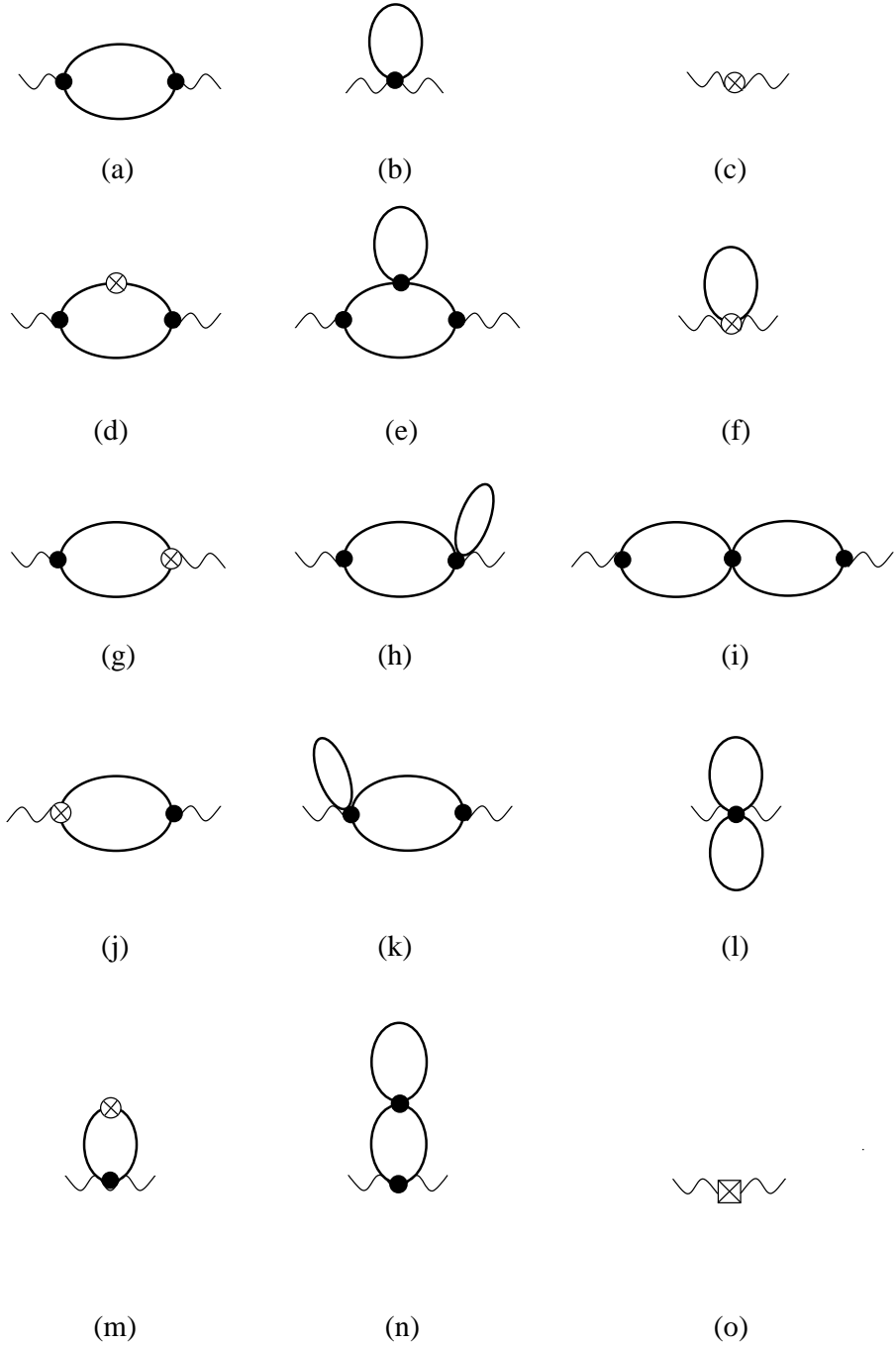,width=12cm} 
\end{center}
\caption[VV diagrams.]
{\label{VVdiag} Diagrams for the vector
two-point function.
The crossed circle stands for the $\pfour$ vertex insertion. Wiggly lines
are the external vector currents. Dots are ${\cal O}(p^2)$ vertices
and a square is a $\psix$ vertex. The solid lines are meson propagators.}
\end{figure}

The $\pfour$ contributions come from diagrams (a-c) in Fig. \ref{VVdiag}.
Diagrams (d,e,m) and (n) can be calculated directly or using
wave-function renormalization and mass corrections. We have checked that
both approaches give the same result. As a consequence the final result
contains no three-propagator integrals. They cancel when 
the $\pfour$ result is expressed in terms
of the physical masses using
\be
f(m_{i_0}^2;q^2) = f(m_i^2;q^2) + \sum_j (m_{j0}^2-m_j^2)
\frac{\partial}{\partial m_{j_0}^2} f(m_{i_0}^2;q^2)
\Big |_{m_{i_0}^2 = m_{i}^2},
\ee
where the $m_{i_0}^2$ are the bare masses and
the $ m_{i}^2$ the next-to-leading order
masses. In addition we replace $F_0$ by $F_\pi$ and all masses by their
physical ones in the $\psix$ expression.

There are no one-particle reducible contributions to the vector two-point
functions.

We have performed the following checks
\begin{enumerate}
\item In the isospin and hypercharge case the longitudinal part vanishes.
\item In the SU(3) limit, i.e.
$m_\pi^2 = m_K^2 = m_\eta^2 \to m_{chiral}^2$, all two-point functions are
equal.
\item The SU(3) breaking effect in the form-factors appears
only in second order in the quark masses, i.e. order $(m_s-\hat m )^2$,
as required by the Ademollo-Gatto theorem
\cite{AD}.
\item All divergences with a non-analytical dependence on masses or $q^2$
cancel and the $\ln(4\pi)$ and $\gamma$ terms can be absorbed in the
counter-terms as well. Both of these follow from general
renormalizability theorems.
\item The remaining divergences are in agreement with those from
the general calculation using heat-kernel methods \cite{BCG}
and with the double logarithms of \cite{BCG1}.
\end{enumerate}

The result can be expressed simply in terms of the finite functions
\ba
\mu_i&=&\frac{m_i^2}{32\pi^2}\ln\frac{m_i^2}{\mu^2}\;,
\nonumber\\
B_V^{ij}(q^2) &=&
\frac{1}{q^2}\left\{ {B}_{22}(m_i^2,m_j^2,q^2)-\frac{\lambda_0}{64\pi^2}
\left(m_i^2+m_j^2-\frac{q^2}{3}\right)+\frac{1}{2}(\mu_i+\mu_j)
\right\}\,,
\nonumber\\
\ea
for $i,j=\pi,K,\eta$. The definitions of $B_{22}(m_i^2,m_j^2,q^2)$
and $\lambda_0$ can be found in
App.~\ref{oneloopint}. Notice that $B_V^{ii}(q^2)$ 
is regular at $q^2=0$ for $i=j$.

For the isospin transverse part we find
\ba
\lefteqn{\Pi_{V\pi}^{(1)}(q^2) = -4 (2 H_1^r + L_{10}^r) 
-8 {B}_V^{\pi\pi}(q^2)-4 {B}_V^{KK}(q^2) }&&
\nonumber\\&&+
\frac{1}{F^2_\pi} \bigg\{
-16 \bigg( 2 q^2 {B}_V^{\pi\pi}(q^2) + q^2 {B}_V^{KK}(q^2) \bigg) L_9^r
+16 q^2 (2 \mu_\pi + \mu_K) (L_9^r+L_{10}^r)
\nonumber\\&&+
4 q^2\bigg( 2 {B}_V^{\pi\pi}(q^2) + {B}_V^{KK}(q^2) \bigg)^2
\bigg\}
-32 m_\pi^2 C^r_{61} - 32  (m_\pi^2+2 m_K^2) C^r_{62} -8q^2 C^r_{93}\,.\nonumber\\
\ea

The hypercharge transverse part is given by
\ba
\lefteqn{
\Pi_{V\eta}^{(1)}(q^2) = 
-4 (2 H_1^r + L_{10}^r) - 12 {B}_V^{KK}(q^2) }&& 
\nonumber\\&&+
\frac{1}{F^2_\pi} \bigg\{-48 q^2{B}_V^{KK}(q^2) L_9^r 
 + 48 \mu_K (L_9^r+L_{10}^r)
 +36 q^2 ({B}_V^{KK}(q^2))^2
\bigg\}
\nonumber\\&&-
32  m_\eta^2 C^r_{61} - 32 (m_\pi^2+2 m_K^2) C^r_{62}-8 q^2 C^r_{93}\;.
\ea
The longitudinal part vanishes for the above two.
These results agree with those obtained in \cite{KG1} when the differences
in subtraction schemes are taken into account.

The expressions for the kaon two-point functions are new and
are somewhat longer. The transverse part is given by
\ba
\lefteqn{
\Pi_{VK}^{(1)}(q^2) = -4 (2 H_1^r + L_{10}^r) - 6
({B}_V^{\pi K}(q^2)+ {B}_V^{\eta K}(q^2)) }&&
\nonumber\\&&+
\frac{1}{F^2_\pi} \bigg\{ 
-\frac{8}{q^2} (m_K^2-m_\pi^2) (3\mu_\pi-2\mu_K-\mu_\eta) L_5^r 
+12  (\mu_\pi +2\mu_K+\mu_\eta) (L_9^r+L_{10}^r)
\nonumber\\&&-
24 q^2 \bigg({B}_V^{\pi K}(q^2)+{B}_V^{\eta K}(q^2)\bigg) L_9^r
-\frac{3}{4q^2} (5 \mu_\pi \mu_\pi -4 \mu_K \mu_K
-3 \mu_\eta \mu_\eta -4 \mu_\pi \mu_K 
\nonumber\\&&-
6 \mu_\pi \mu_\eta+12 \mu_K \mu_\eta)
+9q^2 \bigg({B}_V^{\pi K}(q^2) + {B}_V^{\eta K}(q^2) \bigg)^2
\bigg\}
-\frac{32}{q^2} (m_\pi^2-m_K^2)^2 C^r_{38}
\nonumber\\&&-
32  m_K^2 C^r_{61}-32 (m_\pi^2 +2 m_K^2)C^r_{62}
-\frac{8}{q^2}(m_\pi^2-m_K^2)^2 C^r_{91} - 8 q^2C^r_{93}\;.
\ea
This two-point function has also a longitudinal part
\ba
\lefteqn{
\Pi_{VK}^{(0)}(q^2) =  
\frac{3}{2 q^4} \bigg ( (m_K^2-m_\pi^2)^2 \bar{J}^{\pi K}(q^2) + (m_K^2-m_\eta^2)^2
\bar{J}^{\eta K}(q^2) \bigg ) }&&
\nonumber\\&&+
\frac{1}{q^2 F^2_\pi} \bigg\{ 
\bigg ( 8(m_K^2-m_\pi^2) (3 \mu_\pi-2 \mu_K-\mu_\eta)
+12(m_K^2 -m_\pi^2)^2 \bar{J}^{\pi K}(q^2) 
\nonumber\\&&+
12(m_K^2-m_\eta^2)^2
\bar{J}^{\eta K}(q^2) \bigg ) L_5^r
+\frac{3}{4} ( 5 \mu_\pi \mu_\pi -4 \mu_\pi \mu_K 
 -6 \mu_\pi \mu_\eta -4 \mu_K \mu_K
\nonumber\\&&+
12 \mu_K \mu_\eta -3 \mu_\eta \mu_\eta ) -
 \frac{3}{4} ( -5\mu_\pi +2 \mu_K +3 \mu_\eta)(m_K^2 -m_\pi^2) 
 \bar{J}^{\pi K}(q^2)
\nonumber\\&&-
\frac{9}{4} ( \mu_\pi -2 \mu_K + \mu_\eta)
 (m_K^2-m_\eta^2)\bar{J}^{ \eta K}(q^2)
+\frac{3}{2} \left((m_K^2 -m_\pi^2) \bar{J}^{\pi K}(q^2)\right)^2
\nonumber\\&& 
-\frac{9}{16} \bigg((m_K^2 -m_\pi^2)\bar{J}^{\pi K}(q^2)
+ (m_K^2-m_\eta^2)\bar{J}^{ \eta K}(q^2) \bigg)^2
\nonumber\\&&+
\frac{1}{q^2} \bigg( -\frac{3}{8} (m_\pi^2+m_K^2) 
\left((m_K^2 -m_\pi^2) \bar{J}^{\pi K}(q^2) \right)^2
\nonumber\\&&+
\frac{3}{4} (m_\pi^2+m_K^2) (m_K^2 -m_\pi^2) (m_K^2-m_\eta^2) 
 \bar{J}^{\pi K}(q^2) \bar{J}^{ \eta K}(q^2)
\nonumber\\&&-
(\frac{3}{8} m_\pi^2-\frac{13}{8} m_K^2)\left((m_K^2-m_\eta^2) 
  \bar{J}^{ \eta K}(q^2)\right)^2 \bigg)
\nonumber\\&&-
\frac{9}{16q^4} \bigg( (m_K^2 - m_\pi^2)^2 \bar{J}^{\pi K}(q^2)
+(m_K^2-m_\eta^2)^2 \bar{J}^{\eta K}(q^2) \bigg)^2 \bigg\}
\nonumber\\&&
+\frac{8}{q^2}(m_\pi^2-m_K^2)^2 (4 C^r_{38}+C^r_{91})\;,
\label{PiVK}
\ea
Notice that 
the Ademollo-Gatto theorem \cite{AD} is explicitly satisfied.

All divergences have been absorbed in the coefficients of the
$\psix$ Lagrangian
by setting
\ba
\mu^{4\epsilon}(4 C_{38}+C_{91}) &=&4 C^r_{38}+C^r_{91}
+\frac{5}{32F_0^2}\lambda_2
  -\frac{5}{3F_0^2}\lambda_1 L^r_5\;,
\nonumber\\
\mu^{4\epsilon}C_{61} &=& C^r_{61}
 -\frac{3}{8F_0^2}\lambda_1(L^r_9+L^r_{10})\;,
\nonumber\\
\mu^{4\epsilon}C_{62} &=& C^r_{62}
 -\frac{1}{8F_0^2}\lambda_1(L^r_9+L^r_{10})\;,
\nonumber\\
\mu^{4\epsilon}C_{93} &=& C^r_{93}
 -\frac{1}{32F_0^2}\lambda_2 +\frac{1}{2F_0^2}\lambda_1 L^r_9\;,
\ea
which agrees with the calculation of \cite{BCG}.

\section{Masses and Decay Constants}
\label{masses}

\subsection{Masses}

\begin{figure}
\begin{center}
\epsfig{file=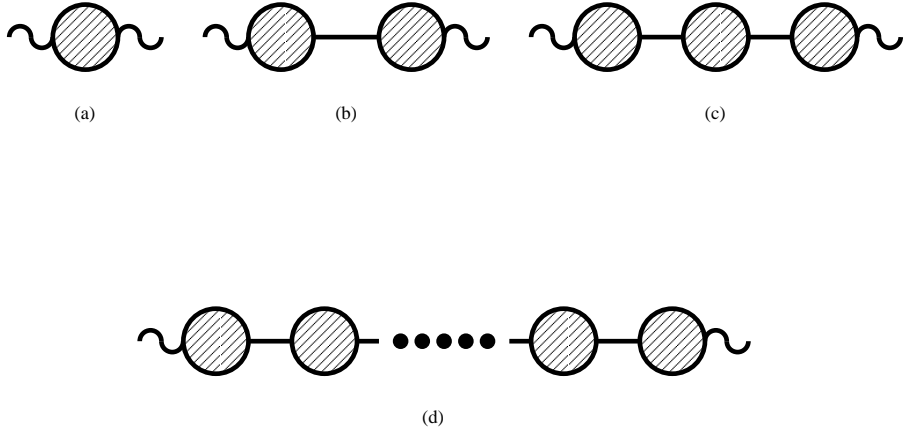,width=12cm}
\end{center}
\caption{\label{figAA1PI} The diagrams contributing to the
axial-vector two-point function. The filled circles indicate
the one-particle-irreducible diagrams. Solid lines are pseudoscalar meson
propagators and the wiggly lines indicate insertions of an
axial-vector current. For the inverse propagator the wiggly lines
are meson legs and for the decay constant the right wiggly line is a meson
leg while the left remains an axial current.}
\end{figure}

\begin{figure}
\begin{center}
\epsfig{file=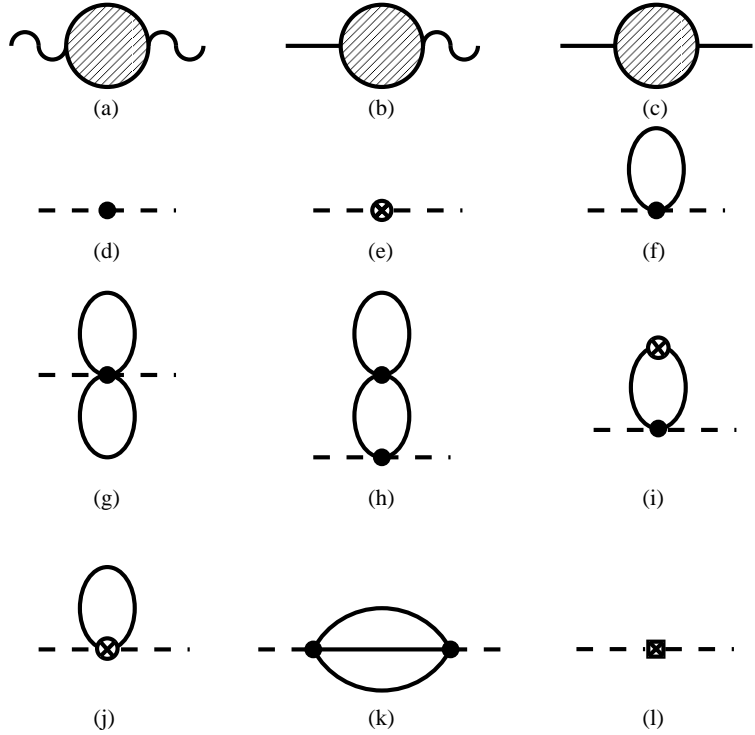,width=10cm}
\end{center}
\caption{\label{figAA} The set of diagrams contributing to the
1PI quantities. (a) axial-vector--axial-vector (b) axial-vector--pseudoscalar
(c) pseudoscalar--pseudoscalar. (d)-(l) the respective diagrams when the
dashed lines are replaced with the external legs of (a), (b) or (c).
A line is a meson propagator, a wiggly line an external source, a dot a vertex of
${\cal O}(p^2)$, a crossed circle a vertex of $\pfour$ and a crossed
box a vertex of $\psix$.}
\end{figure}

The definition of mass is the 
position of the pole in a two-point
Green function $G(p^2,m^2)$
that contains the relevant particle as a possible
intermediate state.
The axial-vector two-point function is a suitable candidate to obtain the 
masses for the pseudoscalar mesons. 
The general structure of this two-point function is shown in
terms of one-particle-irreducible (1PI) diagrams in
Fig. \ref{figAA1PI}.
However the meson-propagator or the two-point function of meson fields
itself is a simpler set with the same pole. We denote the sum of
1PI graphs by $i\Pi(p^2,m_{i0}^2,F_0)$.
The set of diagrams contributing to $\Pi(p^2,m_{i0}^2,F_0)$ is depicted in
Fig. \ref{figAA}(c).
The propagator is
\ba
\label{defmass}
G(p^2,m_{i0}^2,F_0) &=& \frac{i}{p^2-m_0^2}
\sum_{n=0}^\infty\left(
i\Pi(p^2,m_{i0}^2,F_0)\frac{i}{p^2-m_0^2}\right)^n\nonumber\\
&=&\frac{i}{p^2-m_0^2+\Pi(p^2,m_{i0}^2,F_0)}\,,
\ea
where $m_0^2$ stands for the lowest order mass and $m_{i0}^2$
collectively denotes the various lowest order masses.
The physical mass is given by the zero of the denominator 
once the external legs are on mass-shell
\be
m_{phys}^2-m_0^2+\Pi(m_{phys}^2,m_{i0}^2,F_0) = 0\,.
\ee
We replace the masses $m_{i0}$ by their physical masses and $F_0$
by $F_\pi$.
It is sufficient to use the NLO formulae for these in $\Pi(p^2,m_{i0}^2,F_0)$
since it is already of $\pfour$.
\noindent
This leads to
\be
m_{phys}^2 = m_0^2 + (m^2)^{(4)} + (m^2)^{(6)}\,,
\label{massesphys}
\ee
where the bare masses appear only at the leading order and superscripts refer 
to the chiral order.

The resulting formulae for the pion, kaon and eta masses are
gathered in
App. \ref{appmass}. The formulae for the 
pion and eta mass agree
with those of \cite{KG4} (the explicit formulae only appear in the preprint
version) in the way described earlier, see footnote \ref{footnote1},
while the kaon result is new.

Notice that the precise expression for the $\psix$
is dependent on the choice of the $\pfour$ expression\footnote{
This is why our expressions have some differences with those of \cite{KG4}
even after correcting for the renormalization scheme.}, using the 
Gell-Mann-Okubo relation at $\pfour$, produces differences at $\psix$.

The masses depend on seven combinations of the $\psix$ constants. All
the relevant checks described in Sect. \ref{vector} were done.

\subsection{Decay Constants}

The pseudoscalar decay constants, $F_a$, are defined by
\be
\label{defdecay}
\langle 0| A^a(0)_\mu | \phi^a(p)\rangle
= i\sqrt{2}p_\mu F_a\,.
\ee
They can be obtained directly from the definition or by the residue
of the pole in the axial-vector two-point function.
We have calculated them first using their definition and verified that
the calculation via the two-point function yields the same results.

This calculation involves the use of the expression for $\Pi(p^2,m_{i0}^2,F_0)$
obtained earlier to get the wavefunction renormalization in addition
to those diagrams of Fig. \ref{figAA}(b) for the matrix element itself.

We can then write the results in the form
\be
F_a = F_0 \left(1 + \bar F_a^{(4)} + \bar F_a^{(6)}\right)\,.
\ee
Similarly to the masses, the precise form of 
$\bar F_a^{(6)}$ depends on the choice of
$\bar F_a^{(4)}$. For $F_\pi$ and $F_\eta$ we have checked the
double logarithms with those presented 
in \cite{KG4} and the result for $F_K$ is new.

The explicit formulae are rather long and can be found in App. \ref{appdecay}.
The relevant checks described in Sect. \ref{vector} were performed.

\section{The Axial-Vector Two-Point Functions}
\label{axial}

The axial-vector two-point functions to lowest order are quite simple
and all three reduce to
\be
\Pi^{Aa}_{\mu\nu}(q) = 2 F_0^2 g_{\mu\nu}
-\frac{2F_0^2}{q^2-m^2_{0a}} q_\mu q_\nu\,.
\ee
The NLO corrections only introduce minor changes. The decay constants change
to $F_a$, the masses to the physical ones
and there is an additional contribution from the
$\pfour$ constants to $\Pi^{(1)}_{Aa}(q^2) = 4L_{10}^r-8H_1^r$.
In fact a very large part of the $\psix$ corrections is of a similar nature.
We thus define
\be
\Pi^{Aa}_{\mu\nu}(q) = 2 F^2_a g_{\mu\nu}-\frac{2F^2_a}{q^2-m^2_{a}}q_\mu q_\nu
+(q_\mu q_\nu-g_{\mu\nu}q^2)\hat\Pi^{(1)}_{Aa}(q^2)
+q_\mu q_\nu\hat\Pi^{(0)}_{Aa}(q^2)\,.
\label{PiA}
\ee
\noindent
The function $\hat\Pi^{(1)}_{Aa}(q^2)$ can be fully calculated from
diagram (a) of Fig. \ref{figAA1PI}. These are depicted in more detail
in Fig. \ref{figAA}(d-l) and discussed in App. \ref{transverse}.

All the diagrams in Fig. \ref{figAA1PI} contribute to
$\hat\Pi^{(0)}_{Aa}(q^2)$ even though most of their contents
actually go into the redefinitions of the respective decay constants
and masses. The full result is given in App. \ref{longitudinal}. The results 
fulfill the same checks as in Sect. \ref{vector}.
We call $\hat\Pi^{(0)}_{Aa}(q^2)$ and $\hat\Pi^{(1)}_{Aa}(q^2)$
the longitudinal and transverse remainder respectively.

\section{Estimates of some $\psix$ constants}
\label{estimates}

In this section we estimate some of the ${\cal O}(p^6)$ 
constants that appear in the results. We assume saturation 
by the lightest vector, axial-vector and scalar mesons, extending the formalism
used in \cite{EG} to the present case.

For the spin-1 
mesons we use the realization where the vector contribution to the chiral 
Lagrangian starts at ${\cal O}(p^6)$. Keeping only the relevant terms for our 
calculation we have
\be
{\cal L}_V  =  -\frac{1}{4} \langle V_{\mu \nu} V^{\mu \nu} -
 2 M^2_V V_\mu V^\mu \rangle 
 - \frac{f_V}{2 \sqrt{2}} \langle V_{\mu \nu} f^{\mu \nu}_{+} \rangle 
+\ldots\,,
\ee
where
\ba
V_{\mu \nu}  =  \nabla_\mu V_\nu - \nabla_\nu V_\mu \,,& &
f^{\mu \nu}_{\pm}  =  
u (v_{\mu \nu}-a_{\mu \nu})  u^\dag 
\pm u^\dag (v_{\mu \nu}+a_{\mu \nu}) u \nonumber \,,
\ea
\noindent 
and the same holds for the axial-vector with the label change $V \leftrightarrow A$ and 
$f_+\leftrightarrow f_-$. $V_\mu$ and $A_\mu$ are three-by-three matrices in
flavour space and describe the full vector and axial-vector nonets, thus we assume
a nonet symmetry throughout the rest of this section.
The rest of the notations was already presented in Sect. \ref{chpt}. 

After integrating out the vectors the terms contributing to 
the two-point Green functions at \psix\ are
\be
{\cal L}_{V,A} = \frac{f_V^2}{4 M^2_V} \langle f^{\mu \nu}_{+} 
\nabla_\mu \nabla_\lambda f^{\lambda \nu}_{+} \rangle +
\frac{f_A^2}{4 M^2_A} \langle f^{\mu \nu}_{-} 
\nabla_\mu \nabla_\lambda f^{\lambda \nu}_{-} \rangle \, .
\label{LVA}
\ee

In the scalar case, the Lagrangian reads
\ba
{\cal L}_S &=& \frac{1}{2} \langle \nabla^\mu S \nabla_\mu S 
 - M^2_S S^2 \rangle  
 + c_d \langle Su^\mu u_\mu \rangle + c_m \langle S \chi_+ \rangle 
 + \frac{d_m}{2} \langle S^2 \chi_+ \rangle 
\nonumber\\&&
+ c_\gamma \langle S f_{+\mu \nu} f_+^{\mu \nu} \rangle
+ c^\prime_\gamma \langle S f_{-\mu \nu} f_-^{\mu \nu} \rangle\, . 
\ea
After integrating out the scalars, the ${\cal O}(p^6)$ contribution
we are interested in comes from the terms
\be
{\cal L}_S = \frac{-c^2_m}{2 M^4_S} \langle \chi_+ \nabla^2 \chi_+ \rangle 
+ \frac{d_m c_m^2}{2 M^4_S} \langle \chi_+^3 \rangle 
+ \frac{c_d c_m d_m}{M^4_S} \langle \chi_+^2 u_\mu u^\mu \rangle 
+ \frac{c_m c_\gamma}{M^2_S} \langle \chi_+ f_{+\mu \nu} f_+^{\mu \nu} \rangle
\, ,
\ee
obtained after the shift of the vacuum expectation value and using 
the equation of motion for the scalars. Note that only the relevant 
terms are written and, as in the vector and axial case,
a full nonet of scalars is assumed in $S$. 

As input parameters we use 
\begin{eqnarray}
\label{resonanceinput}
M_V = M_\rho = 0.77 \mbox{ GeV}, & & f_V = 0.20,\quad
M_A = M_{a_1} = 1.23 \mbox{ GeV}, \quad f_A = 0.10,
\nonumber\\
M_S = 0.93 \mbox{ GeV}, & & c_m = 0.042 \mbox{ GeV}, \quad
c_d = 0.032 \mbox{ GeV},\nonumber\\
c_\gamma = 19\cdot10^{-3}  \mbox{ GeV}^{-1},& & 
c^\prime_\gamma \sim c_\gamma \quad 
\mbox{and}\quad 
d_m = -2.4\,.
\end{eqnarray}
$M_S$ and 
$d_m$ are obtained from the 
masses of the scalars $K^*_0(1430)$ and $a_0(980)$. The value 
$c_\gamma$ is obtained from 
$\Gamma(f_0 \rightarrow \gamma \gamma)=0.56\pm0.11$~keV and is compatible also
with $\Gamma(a_0\to\gamma\gamma)= 0.30\pm0.10$~keV.
The values of $c_m$ and $c_d$ are obtained forcing the saturation of 
some of the ${\cal O}(p^4)$ constants by the scalars \cite{EG}
and are compatible with those obtained in $\Gamma(a_0\to\pi\eta)$.
$c^\prime_\gamma$ value cannot be determined from data at present,
we assume a value similar to $c_\gamma$.

Using the notation of \cite{p6} for the ${\cal O}(p^6)$ terms, 
the spin-1 Lagrangian yields
\ba
C^r_{87}  \sim  \frac{1}{8} \left( \frac{f^2_V}{M^2_V} 
-\frac{f^2_A}{M^2_A} \right)  \sim  7.6 \cdot 10^{-3} \, \mbox{ GeV}^{-2} \,, 
\\
C^r_{93}  \sim  - \frac{f^2_V}{4 M^2_V}  \sim 
- 17 \cdot 10^{-3} \, \mbox{ GeV}^{-2} \, ,
\label{Cr93}
\ea
and the scalar Lagrangian estimates
\ba
C^r_{14}  \sim  \frac{c_d c_m d_m}{M^4_S} \sim  -4.3  \cdot 10^{-3}\, 
\mbox{ GeV}^{-2}\,, \\
C^r_{19}  \sim  \frac{d_m c^2_m}{2 M^4_S}  \sim  -2.8  \cdot 10^{-3}\, 
\mbox{ GeV}^{-2}\,, \\
C^r_{38}  \sim  \frac{c^2_m}{2 M^4_S}  \sim 1.2 \cdot 10^{-3}\, 
\mbox{ GeV}^{-2}\,, \label{Cr38} \\
C^r_{61} \sim \frac{c_m c_\gamma}{M^2_S} \sim 1.9 \cdot 10^{-3}\, 
\mbox{ GeV}^{-2}\,,\\
C^r_{80} \sim \frac{c_m c^\prime_\gamma}{M^2_S} \sim 1.9 \cdot 10^{-3}\, 
\mbox{ GeV}^{-2}\,.
\ea

We stress that the goal of this section is to roughly estimate
the values of the $\psix$ constants.
A real determination would imply the use of chiral sum rules or other 
processes to fix them. It is 
worth mentioning that the result in Eq. (\ref{Cr93}) is the same if we 
use an 
antisymmetric formalism for the vector Lagrangian, and is in agreement
with the 
result extracted from the experimental data using sum rules 
for the vector-vector two-point functions \cite{KG2}. The generalization 
to the three flavours introduces a new relation
of $\psix$ constants,  $4 C^r_{38}+ C^r_{91}$, due to the 
explicit chiral symmetry breaking in the kaon Green function.

Finally, we remark that the precise value of the constants can have an 
important variation depending of the input values in 
Eq. (\ref{resonanceinput}). Consequently, although the values cited 
in this section are used for the numerical results, 
with the understanding that the other counter-terms are set to zero, 
we have to keep in mind that these values could overestimate the 
physical ones. The latter is especially true for $d_m$
since the $K_0$ and $a_0$ mass difference appears unnaturally large.

\section{Some numerical results}
\label{numerics}

We defer a more accurate comparison with experimental data to \cite{sumrules},
but we would like to present some results using our explicit expressions.
We use the  values for the $C_i^r$
obtained in the previous section and
two \emph{sets} of the \pfour\ constants. They only differ in the values of
$L_1^r$, $L_2^r$ and $L_3^r$.
Set~A is obtained from the fit of the unitarized $K_{l4}$ calculation
while set~B refers to $K_{l4}$ and $\pi \pi$ data at one 
loop accuracy 
\cite{dispersivekl4}. We give both sets to show
an example of the variation with the \pfour\ constants.
We do not show results for varying the other $L_i^r$ but this
results in a similar variation in size of the \psix\ results.
The explicit values we use, at $\mu=0.77~GeV$, are
\begin{eqnarray}
L_1^r & = & 0.37 \cdot 10^{-3}, \quad  L_2^r = 1.35\cdot 10^{-3}, 
\quad L_3^r = -3.5\cdot 10^{-3}\quad\mbox{{\bf set A}} \nonumber \\
L_1^r & = & 0.60 \cdot 10^{-3}, \quad  L_2^r = 1.50\cdot 10^{-3}, 
\quad L_3^r = -3.3\cdot 10^{-3}\quad\mbox{{\bf set B}} \nonumber \\
L_4^r & = & -0.3 \cdot 10^{-3}, \quad  L_5^r = 1.4\cdot 10^{-3}, 
\quad L_6^r = -0.2\cdot 10^{-3}, \nonumber \\
L_7^r & = & -0.4\cdot 10^{-3},\quad  L_8^r = 0.9 \cdot 10^{-3}, 
\quad  L_9^r = 6.9 \cdot 10^{-3}, \nonumber \\
L_{10}^r & = & 1/4  (f_V^2-f_A^2) =  -7.5 \cdot 10^{-3},\nonumber\\
H_1^r & = & 1/8     (f_V^2+f_A^2) = 6.25 \cdot 10^{-3}. 
 \nonumber \\
\end{eqnarray}
For $H_1^r$, which can not be obtained experimentally, we take the value 
from the Meson Saturation Model. Because the vector contribution should 
cancel for the axial-vector two-point function, we use the same model value
for $L^r_{10}$.

The rest of the quantities we use are
\be
m_\pi = 0.135 \, {\rm GeV}, \quad
m_K = 0.495 \, {\rm GeV},\quad 
m_\eta = 0.548 \, {\rm GeV} ~~\mbox{and}~~
F_\pi = 0.0924 \, {\rm GeV}.
\ee
They seem reasonable averages of the various isospin related ones.

\subsection{The Vector Two-Point Functions}

\begin{figure}
\begin{center}
\epsfig{file=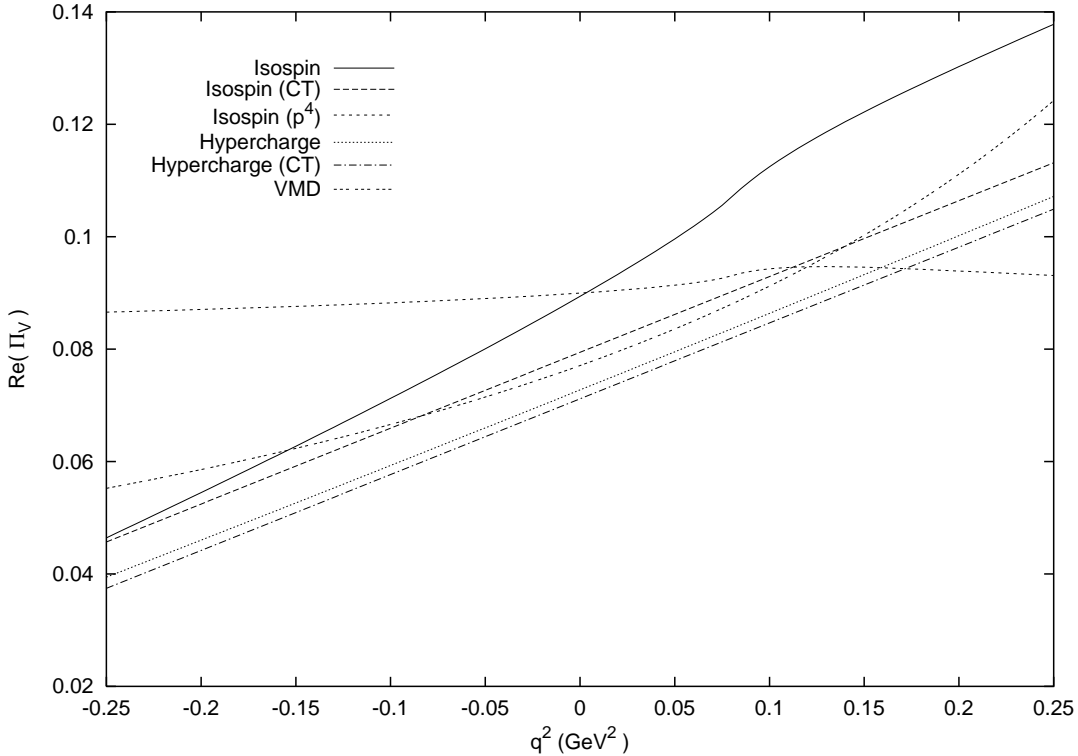,width=10cm,angle=-90}
\end{center}
\caption{\label{figVVnumerics1} Different contributions to the real part
of the vector-vector two-point function in the isospin and
the hypercharge case. The label CT indicates the contribution
from only the counter-terms $L_i^r$ at \pfour\
and $C_i^r$ at \psix. $p^4$ gives the \pfour\
result. VMD indicates the vector model of Eq.~(\ref{VMD}).}
\end{figure}

\begin{figure}
\begin{center}
\epsfig{file=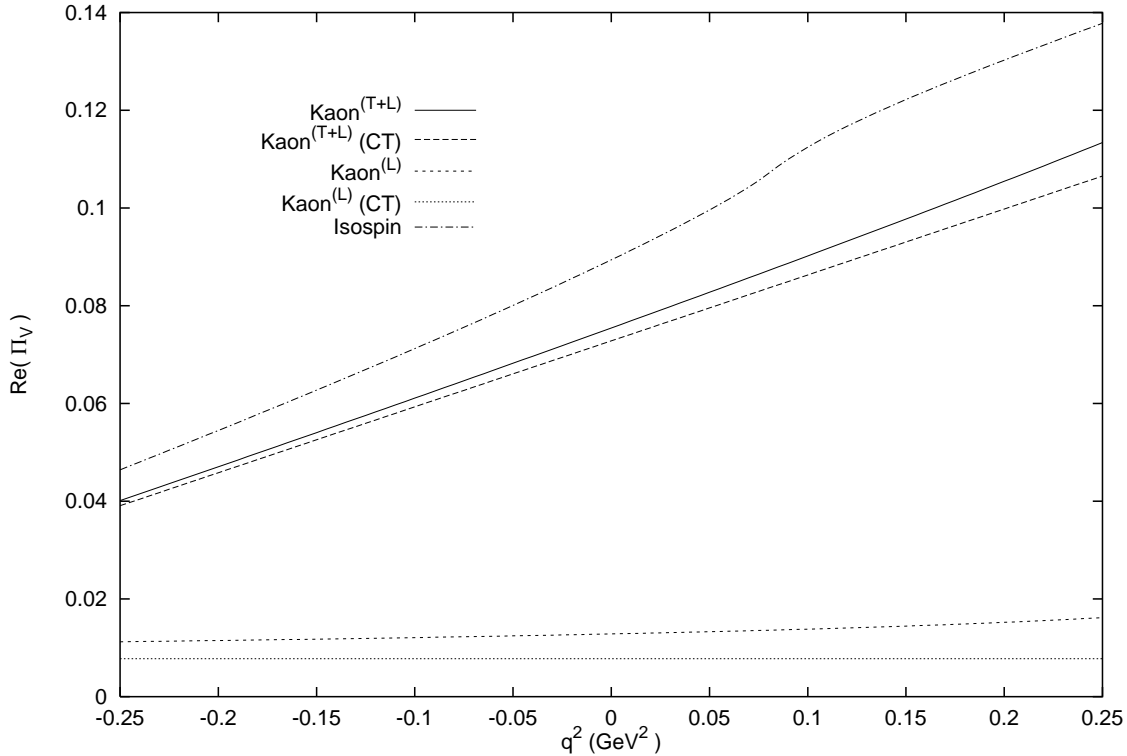,width=10cm,angle=-90}
\end{center}
\caption{\label{figVVnumerics2} Different contributions to the real
part of the kaon vector-vector two-point
function.
The superscripts L (T) refers to the longitudinal (transverse) part. 
For comparison we also plot the isospin case. (CT) as in 
Fig.~\ref{figVVnumerics1}.}
\end{figure}

In Fig.~\ref{figVVnumerics1} and Fig.~\ref{figVVnumerics2} we plot 
the real part of the three vector-vector 
two-point functions choosing set A inputs. 

In all the three cases the slopes
are given mainly by the \psix\ constants as estimated above.
We have shown these contributions in the curves labeled (CT). 
Essentially  --- and with exception of
$C^r_{93}$ --- the main effect of varying the input parameters
is to shift the plots vertically. We see that the 
loop effects are larger in the isospin case and smaller for both,
the hypercharge and kaon. 
In the chiral limit all three cases reduce to the same, and differences are 
related to the breaking of the $SU(3)$ symmetry. For the isospin, 
the two-pion channel produces the notable difference with the counter-term 
contribution, while for the hypercharge and kaon the smaller difference
is explained by the larger masses in the loops and some
explicit breaking of the symmetry through the quark masses in the counter-term
contributions.

In Fig.~\ref{figVVnumerics1} we also plotted the case with a 
complete saturation by the vector meson (VMD)
\be
Re( \, \Pi_V \, ) 
\sim \frac{2 f_V^2 (1-\frac{q2}{M^2_V})}{(1- \frac{q^2}{M^2_V})^2 
+ \frac{\Gamma^2}{M^2_V}},\,
\label{VMD}
\ee
\noindent
with $\Gamma = 0.150 \, {\rm GeV}$. 
The conclusion is that models with only vectors      
explain the main part of the two-point function, however an
important contribution coming from the two-pion intermediate
states is present. 
The curve including only the counter terms ---isospin (CT)---
can also be obtained with the first two terms of the expansion in the previous
formula with $\Gamma=0$ and considering the 
tiny modification due to the scalars.

\subsection{Masses and Decay Constants}

We continue our discussion with the masses and decay constants.
We have summarized our numerical results in Table \ref{massedecay}
using the values
for the \pfour\ constants quoted above.
As one sees in columns three to six, 
both masses and decay constants have substantial
\psix\  loop contributions. 
In addition the pure polynomial piece at \psix\
tends to have the opposite sign and is very large using
our model dependent estimates. This only reinforces the
statement in Sect. \ref{estimates} of the lack of knowledge
in the scalar sector. The terms containing $d_m$ are the only ones
contributing in this subsection and seem severely overestimated
even though they are of a size expected by 
\emph{naive} dimensional analysis.

\begin{table}
\begin{center}
\begin{tabular}{c|cccccc}
 & \pfour  & set A & set A  & set A
 & set B & $C_i^r$\\
\hline
$\mu$ (GeV) & 0.77 & 0.77 & 0.5 & 1.0 & 0.77 & $-$\\
$F_\pi/F_0$ &
0.068 & $-$0.101 & $-$0.066 & $-$0.100 & $-$0.172 &$-$0.001\\
$(F_\pi/F_0)$   &    &    (0.013)& & & ($-$0.050) & \\
$F_K/F_\pi$ &
0.216 & 0.055 & $-$0.023 & 0.100 &0.035 & $-0.450$\\
($F_K/F_\pi$) & &  (0.08) & & & (0.06) & \\
$F_\eta/F_\pi$ &
0.312 & 0.092 & $-$0.011 & 0.150 & 0.065 & $-0.600$\\
$m_\pi^2/m_{\pi phy}^2$ &
$-$0.039 & 0.214 & 0.132 & 0.238 & 0.355 & $-$0.003\\
$m_K^2/m_{K phy}^2$ &
$-$0.003 & 0.241 & 0.246 & 0.194 & 0.423 & $-$0.873
\\
$m_\eta^2/m_{\eta phy}^2$ &
$-$0.045 & 0.312 & 0.234 & 0.273 & 0.521 & $-$2.428
\\
\end{tabular}
\end{center}
\caption{\label{massedecay}
\pfour\ and \psix\ contributions to the masses and decay constants. 
The number in brackets are the extended
double log approximation of Ref.~\cite{BCG1}. Columns three to 
six are the $\psix$ loop contributions only. The contribution from 
the $C_i^r$ is listed separately in the last column.}
\end{table}
In order to have a full presentation of the \psix\ contributions
a refit of all \pfour\ coefficients using the full \psix\ expressions
would be needed. We postpone this till after the main other
processes are also calculated to this order given the dependence of
the \psix\ contributions on $L^r_1$--$L^r_3$.
As an example using set A at $\mu = 0.77~GeV$ otherwise but shifting
$L_5^r$ to $1.0 \cdot 10^{-3}$ reproduces the experimental value of $F_K/F_\pi$
when setting $d_m=0$.

For the decay constants 
the $\psix$ contributions to the ratios
are smaller than the $\pfour$, not including the estimates from
scalar exchange to the \psix\ constants.

To judge the effect of the \psix\ contributions on determining the
quark mass
ratios we use the lowest order, \pfour\ and \psix\ formulae in terms of 
physical quantities to obtain the lowest order masses using 
 Eq. (\ref{massesphys}). This leads to
\begin{eqnarray}
\frac{2m_s}{m_u+m_d} &=& \frac{2m_{0K}^2-m_{0\pi}^2}{m_{0\pi}^2}
= 25.9(p^2);\quad 24.9(p^4);\quad24.1(p^6)\quad\mbox{and}\nonumber\\
\frac{2m_s}{m_u+m_d} &=& \frac{3m_{0\eta}^2-m_{0\pi}^2}{2m_{0\pi}^2}
= 24.2(p^2);\quad24.4(p^4);\quad23.3(p^6)
\end{eqnarray}
using the results from set A at $\mu=0.77~GeV$ and $C^r_i =0$.
These ratios can be compared with 
$2 m_s/(m_u+m_d) = 25.5~ \cite{ximojamin};\, 22.8~ \cite{lattice}$\,, 
extracted from QCD sum rules and lattice calculations respectively.

The emerging conclusions about the convergence of the chiral series 
should be very cautious since a full study includes also
the effect of the \psix\ constants. However, while the corrections calculated
are significant they do not show evidence of
a breakdown of the chiral expansion for the quantities presented here.

\subsection{The Axial-Vector Two-Point Functions}

\begin{figure}[h]
\begin{center}
\epsfig{file=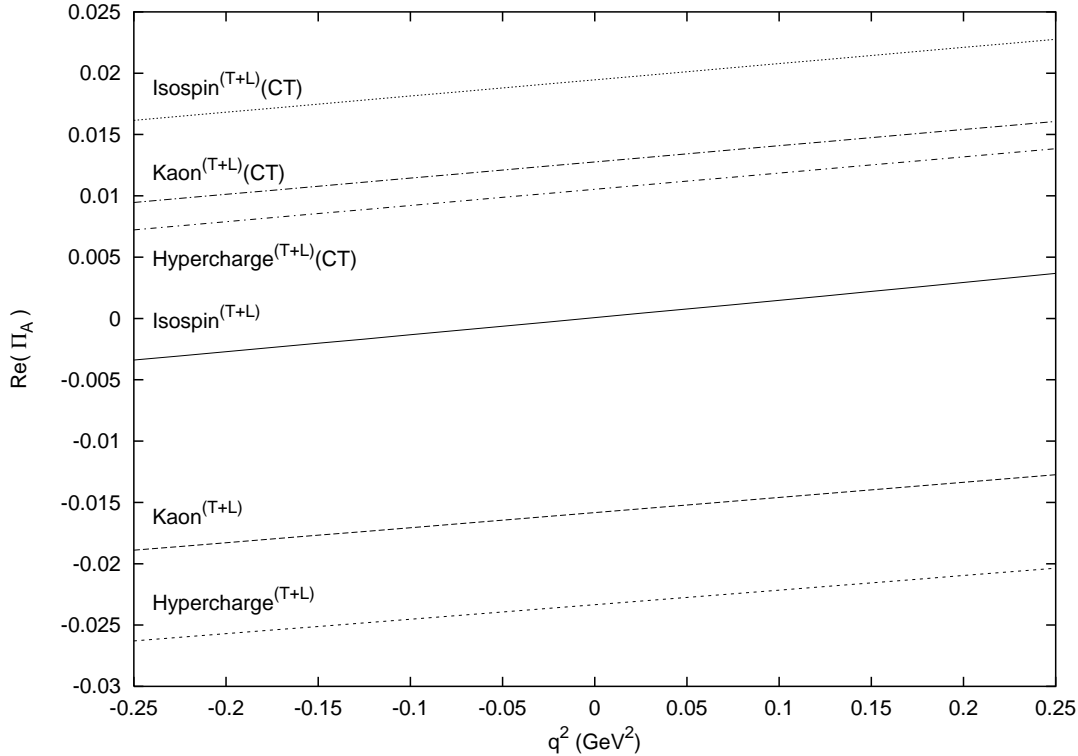,width=10cm,angle=-90}
\end{center}
\caption{\label{figAAnumerics1} Different contributions to the real part
of the remainder axial-axial two-point
function. The superscripts L (T) refers to the 
remainder longitudinal (transverse) part. 
(CT) as in Fig~\ref{figVVnumerics1}.}
\end{figure}

In Fig.~\ref{figAAnumerics1} we plotted the dependence on momenta of the 
real part for the remainders of the
axial-vector two-point functions 
for the three cases under study. Because both the longitudinal and transverse
remainders have poles at $q^2=0$,
we show the combination ${\hat \Pi}_A^{(1)}+{\hat \Pi}_A^{(0)}$. 
A priori we would 
expect a different behaviour for the isospin case due to the three pion 
channel. However there is virtually no effect because
the imaginary part 
is very small in the energy region we are considering in agreement with the 
dominance of the $a_1$ axial meson. For the other two cases even the three 
pseudoscalar channel is far. The curves are thus very linear. 
The vertical shifts are due to the explicit breaking from the quark 
masses. The \psix\ contributions are rather small, the scale in
the plot should be compared with $|2F_\pi^2/q^2|$ which
is larger than 0.07 for the entire region plotted.

\section{Summary and Conclusions}
\label{conclusions}

In this paper we have calculated to NNLO in CHPT the vector and
axial-vector two-point functions in the isospin limit and in the 
complete three flavour basis. 

In the vector-vector case, 
we confirm previous results for the isospin and hypercharge \cite{KG1}.  

For the axial-vector case, 
besides the cancellation of the non-analytic poles, we obtain the same 
double and simple poles that appear with the use of the heat kernel 
expansion \cite{BCG}. 
We also agree with the double logarithms, appearing in previous 
work \cite{KG4}, for the isospin and hypercharge cases. All 
these checks give us confidence about our result.

We have also given expressions to NNLO for the masses and decay constants.
The Lagrangian at \psix\ contains a rather large number of free constants.
We have estimated some of them using a simple resonance estimate
and used this to present some first numerical results for the
two-point functions, masses and decay constants.
We also studied somewhat the $\mu$-dependence of the final result.

Although the corrections are significant they do not show 
evidence of a breakdown of the chiral expansion. For instance, our
estimates of 
the quark mass ratios are in agreement with previous determinations. 
However, the sensitivity to the input values, indicate that 
the $\pfour$ constants need to be refitted 
using the full $\psix$ expressions and that better estimates 
of the $\psix$ constants are necessary.

\section*{Acknowledgments}
We thank Ll. Ametller for a careful reading of the manuscript.
The work of P.~T.
was supported by the Swedish Research Council (NFR).

\newpage
\appendix

\section{Explicit results for the masses and decay constants}

\subsection{Masses}
\label{appmass}

The masses are split as follows
\begin{eqnarray}
m_a^2 = m_{0a}^2 + (m_a^2)^{(4)} + (m_a^2)^{(6)}_{CT}+ (m_a^2)^{(6)}_{loops}\,.
\end{eqnarray}
with $ m_{0a}$ the contribution from the bare masses. In the \psix~ we have
explicitly separated the chiral loop contribution from the
model dependent counter-terms.

In a previous step the masses are obtained in terms of only the bare masses 
(quark masses), we rewrite 
the $\pfour$ contribution with the physical masses implying a modification of 
the $\psix$ terms. In the $\psix$ we can safely replace
bare masses with physical masses. 

For the pion we obtain
\begin{eqnarray}
m_{0\pi}^2 &=& B_0( m_u+m_d) = 2 B_0\hat m\,,
\nonumber\\
\frac{F_\pi^2}{m_\pi^2} (m_\pi^2)^{(4)} &=& 
 8 (m_\pi^2+2 m_K^2)(2L_6^r- L_4^r)
+ 8 m_\pi^2 (2L^r_8-L^r_5) + \mu_\pi-\frac{1}{3} \mu_\eta
\end{eqnarray}
in agreement with \cite{GL}.
The \psix\ contributions are
\begin{eqnarray}
\lefteqn{F_\pi^2 (m_\pi^2)^{(6)}_{CT} =
  - 32m_\pi^6C^r_{12} - 32m_\pi^6C^r_{13} - 16m_\pi^6C^r_{14}
  - 16m_\pi^6
        C^r_{15} - 48m_\pi^6 C^r_{16}
}&&\nonumber\\&&
  - 16m_\pi^ 6C^r_{17} 
  + 48m_\pi^6 C^r_{19} + 80m_\pi^6 C^r_{20}
          + 48m_\pi^6 C^r_{21} + 32m_\pi^6 C^r_{31}
 + 32m_\pi^6 C^r_{32}
\nonumber\\&&
  - 64m_\pi^4m_K^2C^r_{13}
          - 32m_\pi^4m_K^2C^r_{15} + 64m_\pi^4m_K^2C^r_{16}
  + 192m_\pi^4m_K^2C^r_{21} 
\nonumber\\&&
 + 64  m_\pi^4m_K^2C^r_{32} 
 - 64m_\pi^2m_K^4C^r_{16}
  + 64 m_\pi^2m_K^4 C^r_{20}
 + 192 m_\pi^2
         m_K^4 C^r_{21} \,,
\end{eqnarray}
\begin{eqnarray}
\lefteqn{F_\pi^4 (m_\pi^2)^{(6)}_{Loops} =
       \frac{1}{(16\pi^2)^2} \, (  - 527/1296\,m_\pi^6\,\pi^2
    - 3217/1728\,m_\pi^6 - 37/324\,m_\pi^4 \,m_K^2\,\pi^2 }&&
\nonumber\\&&
- 139/216\,m_\pi^4\,m_K^2 - 11/36\,m_\pi^2\,m_K^4\,\pi^2
 - 15/16\,m_\pi^2\,m_K^4
          )
\nonumber\\&&
       + \frac{1}{16\pi^2}\Bigg( \, (  - 2\,m_\pi^4 - 2\,m_\pi^2\,m_K^2 )
\,\mu_\pi
       - 2\,m_\pi^2\,m_K^2 \,\mu_K
       + 4\,m_\pi^6 \,L_1^r \
\nonumber\\&&
       +  ( 74/9\,m_\pi^6 - 16/9\,m_\pi^4\,m_K^2 
 + 104/9\,m_\pi^2\,m_K^4 )\,L_2^r
\nonumber\\&&
       +  ( 56/27\,m_\pi^6 - 16/27\,m_\pi^4\,m_K^2 + 86/27\,m_\pi^2\,m_K^4 )
\,L_3^r \,\Bigg)
       - ( 589/36\,m_\pi^2 + 2\,m_K^2)\,\mu_\pi^2 
\nonumber\\&&
       - 4\,m_\pi^2 \,\mu_\pi\,\mu_K 
       + 5/3\,m_\pi^2 \, \mu_\pi\,\mu_\eta 
       - 112\,m_\pi^4 \, \mu_\pi\,L_1^r 
       - 64\,m_\pi^4 \, \mu_\pi\,L_2^r 
\nonumber\\&&
       - 56\,m_\pi^4 \, \mu_\pi\,L_3^r 
       + ( 144\,m_\pi^4 + 80\,m_\pi^2\,m_K^2 ) \, \mu_\pi\,L_4^r  
       + 96\,m_\pi^4 \, \mu_\pi\,L_5^r  
\nonumber\\&&
       + (  - 256\,m_\pi^4 - 160\,m_\pi^2\,m_K^2 ) \mu_\pi\,L_6^r 
       - 176\,m_\pi^4\mu_\pi\,L_8^r 
       + (  - 2/3\,m_\pi^4\,m_K^{-2} - 7\,m_\pi^2 ) \, \mu_K^2  
\nonumber\\&&
       - 4/3\,m_\pi^2 \mu_K\,\mu_\eta 
       - 128\,m_\pi^2\,m_K^2 \, \mu_K\,L_1^r 
       - 32\,m_\pi^2\,m_K^2 \, \mu_K\,L_2^r 
\nonumber\\&&
       - 40\,m_\pi^2\,m_K^2 \, \mu_K\,L_3^r 
       + ( 16\,m_\pi^4 + 160\,m_\pi^2\,m_K^2 )\,\mu_K\,L_4^r 
       + ( 16\,m_\pi^4 + 32\,m_\pi^2\,m_K^2 )\,\mu_K\,L_5^r 
\nonumber\\&&
       +(  - 32\,m_\pi^4 - 192\,m_\pi^2\,m_K^2 )\, \mu_K\,L_6^r  
       + (  - 32\,m_\pi^4 - 64\,m_\pi^2\,m_K^2 )\,\mu_K\,L_8^r  
\nonumber\\&&
       + (  - 5/9\,m_\pi^4\,m_\eta^{-2} - 29/36\,m_\pi^2 ) \,\mu_\eta^2  
       + ( 32/3\,m_\pi^4 - 128/3\,m_\pi^2\,m_K^2 ) \, \mu_\eta\,L_1^r 
\nonumber\\&&
       + ( 8/3\,m_\pi^4 - 32/3\,m_\pi^2\,m_K^2 ) \, \mu_\eta\,L_2^r 
       + ( 8/3\,m_\pi^4 - 32/3\,m_\pi^2\,m_K^2 ) \, \mu_\eta\,L_3^r  
\nonumber\\&&
       + (  - 16\,m_\pi^4 + 48\,m_\pi^2\,m_K^2 ) \, \mu_\eta\,L_4^r  
       +  (  - 64/9\,m_\pi^4 + 64/9\,m_\pi^2\,m_K^2 ) \, \mu_\eta\,L_5^r
\nonumber\\&&
       +  ( 64/3\,m_\pi^4 - 160/3\,m_\pi^2\,m_K^2 ) \, \mu_\eta\,L_6^r
       + (  - 128/3\,m_\pi^4 + 128/3\,m_\pi^2\,m_K^2 ) \, \mu_\eta\,L_7^r  
\nonumber\\&&
       - 16/3\,m_\pi^4 \, \mu_\eta\,L_8^r 
       +  (  - 128\,m_\pi^6 - 512\,m_\pi^4\,m_K^2 - 512\,m_\pi^2\,m_K^4 )
 \, L_4^{r2} 
\nonumber\\&&
       + (  - 256\,m_\pi^6 - 384\,m_\pi^4\,m_K^2 - 128\,m_\pi^2\,m_K^4 ) \, 
L_4^r\,L_5^r \, 
\nonumber\\&&
       +  ( 512\,m_\pi^6 + 2048\,m_\pi^4\,m_K^2 + 2048\,m_\pi^2\,m_K^4 ) 
\, L_4^r\,L_6^r
\nonumber\\&&
       + ( 512\,m_\pi^6 + 768\,m_\pi^4\,m_K^2 + 256\,m_\pi^2\,m_K^4 ) \, L_4^r\,L_8^r 
\nonumber\\&&
       - 128\,m_\pi^6 \, L_5^{r2} 
       +  ( 512\,m_\pi^6 + 768\,m_\pi^4\,m_K^2 + 256\,m_\pi^2\,m_K^4 ) \,L_5^r\,L_6^r
\nonumber\\&&
       + 512\,m_\pi^6 \, L_5^r\,L_8^r 
       +(  - 512\,m_\pi^6 - 2048\,m_\pi^4\,m_K^2 - 2048\,m_\pi^2\,m_K^4 ) \, L_6^{r2}
\nonumber\\&&
       +  \, (  - 1024\,m_\pi^6 - 1536\,m_\pi^4\,m_K^2 - 512\,m_\pi^2\,m_K^4 ) \, L_6^r\,L_8^r
       - 512\,m_\pi^6 \, L_8^{r2} 
\nonumber\\&&
       +  5/6\,H^F(m_\pi^2,m_\pi^2,m_\pi^2;m_\pi^2)\,m_\pi^4
 - 5/8\,H^F(m_\pi^2,m_K^2,m_K^2;m_\pi^2)\,
         m_\pi^4 
\nonumber\\&&
+ 1/18\,H^F(m_\pi^2,m_\eta^2,m_\eta^2;m_\pi^2)\,m_\pi^4
 + H^F(m_K^2,m_\pi^2,m_K^2;m_\pi^2)\,m_\pi^2\,
         m_K^2
\nonumber\\&&
 - 5/6\,H^F(m_K^2,m_K^2,m_\eta^2;m_\pi^2)\,m_\pi^4 
- 1/8\,H^F(m_\eta^2,m_K^2,m_K^2;m_\pi^2)\,m_\pi^4
\nonumber\\&&
          + 1/2\,H^F(m_\eta^2,m_K^2,m_K^2;m_\pi^2)\,m_\pi^2\,m_K^2 
+ H^F_1(m_\pi^2,m_K^2,m_K^2;m_\pi^2)\,m_\pi^4 
\nonumber\\&&
+ 2
         \,H^F_1(m_K^2,m_K^2,m_\eta^2;m_\pi^2)\,m_\pi^4
 + 3\,H^F_{21}(m_\pi^2,m_\pi^2,m_\pi^2;m_\pi^2)\,m_\pi^4 
\nonumber\\&&
- 3/8\,         H^F_{21}(m_\pi^2,m_K^2,m_K^2;m_\pi^2)\,m_\pi^4 
+ 3\,H^F_{21}(m_K^2,m_\pi^2,m_K^2;m_\pi^2)\,m_\pi^4 
\nonumber\\&&
+ 9/8\, H^F_{21}(m_\eta^2,m_K^2,m_K^2;m_\pi^2)\,m_\pi^4\,.
\end{eqnarray}
\noindent
Where the definitions of $H^F_i(m^2_1,m^2_2,m^2_3;q^2)$ 
appear in App. \ref{loopint}. 

For the kaon we obtain
\begin{eqnarray}
m_{0K}^2 &=& B_0( m_s+\hat m) \,,
\nonumber\\
\frac{F_\pi^2}{m_K^2} (m_K^2)^{(4)} &=& 
 8 (m_\pi^2+2 m_K^2)(2L_6^r- L_4^r)
+ 8 m_K^2 (2L^r_8-L^r_5) + 
\frac{m_\pi^2+3m_\eta^2}{6m_K^2}\mu_\eta
\nonumber\\
\end{eqnarray}
in agreement with \cite{GL}.
The \psix\ contributions are
\begin{eqnarray}
\lefteqn{
F_\pi^2 (m_K^2)^{(6)}_{CT} =
  - 16\,m_\pi^4\,m_K^2\,C^r_{14} - 48\,m_\pi^4\,m_K^2\,C^r_{16} 
+ 16\,m_\pi^4\,m_K^2\,C^r_{17}
          + 48\,m_\pi^4\,m_K^2\,C^r_{19} 
}&&
\nonumber\\&&
+ 48\,m_\pi^4\,m_K^2\,C^r_{20}
 + 48\,m_\pi^4\,m_K^2\,C^r_{21} - 32\,m_\pi^2
         \,m_K^4\,C^r_{13} + 32\,m_\pi^2\,m_K^4\,C^r_{14}
\nonumber\\&&
 - 16\,m_\pi^2\,m_K^4\,C^r_{15} + 64\,m_\pi^2\,m_K^4\,
         C^r_{16} - 32\,m_\pi^2\,m_K^4\,C^r_{17} - 96\,m_\pi^2\,m_K^4\,C^r_{19}
\nonumber\\&&
 - 32\,m_\pi^2\,m_K^4\,C^r_{20} + 
         192\,m_\pi^2\,m_K^4\,C^r_{21} + 32\,m_\pi^2\,m_K^4\,C^r_{32}
 - 32\,m_K^6\,C^r_{12}
 - 64\,m_K^6\,
         C^r_{13}
\nonumber\\&&
 - 32\,m_K^6\,C^r_{14} - 32\,m_K^6\,C^r_{15} - 64\,m_K^6\,C^r_{16} + 96\,m_K^6\,C^r_{19}
          + 128\,m_K^6\,C^r_{20} 
\nonumber\\&&
+ 192\,m_K^6\,C^r_{21} + 32\,m_K^6\,C^r_{31} + 64\,m_K^6\,C^r_{32} \,,
\end{eqnarray}
\begin{eqnarray}
\lefteqn{F_\pi^4 (m_K^2)^{(6)}_{Loops} =
       \frac{1}{(16\pi^2)^2}\,
  \bigg(  - 1/8\,m_\pi^4\,m_K^2\,\pi^2 - 13/24\,m_\pi^4\,m_K^2 - 73/648\,m_\pi^2\, \,\pi^2 }&&
\nonumber\\&&
 - 19/108\,m_\pi^2\,m_K^4 - 763/1296\,m_K^6\,\pi^2 - 4709/1728\,m_K^6\bigg)
       +\frac{1}{16\pi^2}\bigg( - 3/2\,m_K^4 \mu_\pi 
\nonumber\\&&
       + (  - 3/2\,m_\pi^2\,m_K^2 - 3/2\,m_K^4 ) \,\mu_K 
       +  (  - 1/2\,m_\pi^2\,m_K^2 - m_K^4 ) \,\mu_\eta
       + 4\,m_K^6 \,L_1^r \, 
\nonumber\\&&
       + ( 56/9\,m_\pi^4\,m_K^2 - 16/9\,m_\pi^2\,m_K^4 + 122/9\,m_K^6 ) \,L_2^r
       + ( 41/27\,m_\pi^4\,m_K^2
\nonumber\\&&
   - 4/27\,m_\pi^2\,m_K^4 + 89/27\,m_K^6 )\, L_3^r \bigg)
       + (  - 2\,m_\pi^{-2}\,m_K^4 - 27/8\,m_K^2 )\, \mu_\pi^2
       - 3\,m_K^2  \,\mu_\pi\,\mu_K \, 
\nonumber\\&&
       + ( 1/3\,m_\pi^2 - 41/12\,m_K^2 ) \,\mu_\pi\,\mu_\eta  
       - 96\,m_\pi^2\,m_K^2 \,\mu_\pi\,L_1^r  
       - 24\,m_\pi^2\,m_K^2 \,\mu_\pi\,L_2^r  
\nonumber\\&&
       - 30\,m_\pi^2\,m_K^2 \,\mu_\pi\,L_3^r  
       + ( 136\,m_\pi^2\,m_K^2 + 64\,m_K^4 ) \,\mu_\pi\,L_4^r
\nonumber\\&&
       +  ( 24\,m_\pi^2\,m_K^2 + 32\,m_K^4 ) \,\mu_\pi\,L_5^r
       +(  - 176\,m_\pi^2\,m_K^2 - 128\,m_K^4 ) \, \mu_\pi\,L_6^r
\nonumber\\&&
       + (  - 48\,m_\pi^2\,m_K^2 - 64\,m_K^4 ) \,\mu_\pi\,L_8^r  
       + (  - 3/2\,m_\pi^2 - 251/18\,m_K^2 ) \,\mu_K^2 
       - 8/3\,m_K^2 \, \mu_K\,\mu_\eta  
\nonumber\\&&
       - 144\,m_K^4\mu_K\,L_1^r 
       - 72\,m_K^4\mu_K\,L_2^r 
       - 60\,m_K^4\mu_K\,L_3^r 
\nonumber\\&&
       +( 16\,m_\pi^2\,m_K^2 + 160\,m_K^4 ) \, \mu_K\,L_4^r  
       + 64\,m_K^4 \, \mu_K\,L_5^r  
       + (  - 32\,m_\pi^2\,m_K^2 - 256\,m_K^4 ) \, \mu_K\,L_6^r 
\nonumber\\&&
       - 128\,m_K^4 \, \mu_K\,L_8^r  
       + (  - 43/288\,m_\pi^4\,m_\eta^{-2} - 25/32\,m_\pi^2 - 5/9\,m_K^2 ) \,\mu_\eta^2
\nonumber\\&&
       + ( 32/3\,m_\pi^2\,m_K^2 - 128/3\,m_K^4 ) \, \mu_\eta\,L_1^r 
       +( 8/3\,m_\pi^2\,m_K^2 - 32/3\,m_K^4 ) \, \mu_\eta\,L_2^r 
\nonumber\\&&
       +( 14/3\,m_\pi^2\,m_K^2 - 56/3\,m_K^4 ) \, \mu_\eta\,L_3^r \, 
       +(  - 8\,m_\pi^2\,m_K^2 + 64\,m_K^4 ) \, \mu_\eta\,L_4^r \, 
\nonumber\\&&
       +( 16/9\,m_\pi^4 - 8/3\,m_\pi^2\,m_K^2 + 224/9\,m_K^4 ) \,\mu_\eta\,L_5^r
       +( 16/3\,m_\pi^2\,m_K^2 - 256/3\,m_K^4 ) \, \mu_\eta\,L_6^r 
\nonumber\\&&
       +(  - 64/3\,m_\pi^4 + 64\,m_\pi^2\,m_K^2 - 128/3\,m_K^4 )
 \, \mu_\eta\,L_7^r 
       + (  - 32/3\,m_\pi^4 + 112/3\,m_\pi^2\,m_K^2
\nonumber\\&&
 - 64\,m_K^4 ) \, \mu_\eta\,L_8^r 
       + (  - 128\,m_\pi^4\,m_K^2 - 512\,m_\pi^2\,m_K^4 - 512\,m_K^6 ) \, L_4^{^r2}
       +(  - 128\,m_\pi^4\,m_K^2 
\nonumber\\&&
- 256\,m_\pi^2\,m_K^4 - 384\,m_K^6 ) \,  L_4^r\,L_5^r
       + ( 512\,m_\pi^4\,m_K^2 + 2048\,m_\pi^2\,m_K^4 + 2048\,m_K^6 ) \, L_4^r\,L_6^r 
\nonumber\\&&
       +( 128\,m_\pi^4\,m_K^2 + 384\,m_\pi^2\,m_K^4 + 1024\,m_K^6 ) \, L_4^r\,L_8^r
       +(  - 64\,m_\pi^2\,m_K^4 - 64\,m_K^6 ) \, L_5^{^r2}
\nonumber\\&&
       +L_5^r\,L_6^r \, ( 256\,m_\pi^4\,m_K^2 + 512\,m_\pi^2\,m_K^4 + 768\,m_K^6 )
       +L_5^r\,L_8^r \, ( 128\,m_\pi^2\,m_K^4 + 384\,m_K^6 )
\nonumber\\&&
       +(  - 512\,m_\pi^4\,m_K^2 - 2048\,m_\pi^2\,m_K^4 - 2048\,m_K^6 ) \, L_6^{^r2}
       +(  - 256\,m_\pi^4\,m_K^2 
\nonumber\\&&
- 768\,m_\pi^2\,m_K^4 - 2048\,m_K^6 ) \, L_6^r\,L_8^r
       - 512\,m_K^6 \, L_8^{^r2} 
\nonumber\\&&
       +  3/4\,H^F(m_\pi^2,m_\pi^2,m_K^2;m_K^2)\,m_\pi^2\,m_K^2 + 3/8\,H^F(m_\pi^2,m_\pi^2,m_K^2;m_K^2)\,
         m_K^4
\nonumber\\&&
 + 1/4\,H^F(m_\pi^2,m_K^2,m_\eta^2;m_K^2)\,m_K^4 - 3/32\,H^F(m_K^2,m_\pi^2,m_\pi^2;m_K^2)\,
         m_K^4
\nonumber\\&&
 + 9/16\,H^F(m_K^2,m_\pi^2;m_\eta^2;m_K^2)\,m_K^4 
+ 3/4\,H^F(m_K^2,m_K^2,m_K^2;m_K^2)\,
         m_K^4
\nonumber\\&&
 + 181/288\,H^F(m_K^2,m_\eta^2,m_\eta^2;m_K^2)\,m_K^4
 - 3/2\,H^F_1(m_\pi^2,m_\pi^2,m_K^2;m_K^2
         )\,m_K^4 
\nonumber\\&&
- 3/2\,H^F_1(m_K^2,m_\pi^2,m_\eta^2;m_K^2)\,m_K^4 
- 5/4\,H^F_1(m_K^2,m_\eta^2,m_\eta^2;m_K^2)
         \,m_K^4
\nonumber\\&&
 + 9/4\,H^F_{21}(m_\pi^2,m_\pi^2,m_K^2;m_K^2)\,m_K^4
 - 9/32\,H^F_{21}(m_K^2,m_\pi^2;m_\pi^2,
         m_K^2)\,m_K^4
\nonumber\\&&
 + 27/16\,H^F_{21}(m_K^2,m_\pi^2,m_\eta^2;m_K^2)\,m_K^4 
+ 9/4\,H^F_{21}(m_K^2,m_K^2,
         m_K^2;m_K^2)\,m_K^4 
\nonumber\\&&
+ 27/32\,H^F_{21}(m_K^2,m_\eta^2,m_\eta^2;m_K^2)\,m_K^4 \,.
\end{eqnarray}

For the eta we obtain
\begin{eqnarray}
\lefteqn{
m_{0\eta}^2 = B_0\frac{2}{3}( 2m_s+\hat m) \,,
}&&\nonumber\\
\lefteqn{
F_\pi^2 (m_\eta^2)^{(4)} =
       -8  m_\eta^2(  m_\pi^2 + 2m_K^2) L_4^r
       +\frac{8}{3}m_\eta^2 (m_\pi^2-4m_K^2) L_5^r
}&&\nonumber\\&&
       -\frac{16}{3}(  m_\pi^4 -2 m_\pi^2 m_K^2 - 8 m_K^4 ) L_6^r
       +\frac{16}{3} ( 3 m_\pi^4 - 8 m_\pi^2 m_K^2 + 8 m_K^4 ) L_8^r
       -m_\pi^2 \mu_\pi 
\nonumber\\&&
       +  (   \frac{2}{3} m_\pi^2 + 2 m_\eta^2 ) \mu_K
       + \left(\frac{7}{9}m_\pi^2-\frac{16}{9}m_K^2\right)\mu_\eta
       +\frac{128}{3} (m_K^2-m_\pi^2)^2 L^r_7
\end{eqnarray}
in agreement with \cite{GL}.
The \psix\ contributions are
\begin{eqnarray}
\lefteqn{
F_\pi^2 (m_\eta^2)^{(6)}_{CT}=
    32/27\,m_\pi^6\,C^r_{12} - 32/9\,m_\pi^6\,C^r_{13} + 16/3\,m_\pi^6\,C^r_{14} - 16/9\,
         m_\pi^6\,C^r_{15} 
}&&
\nonumber\\&&
+ 16\,m_\pi^6\,C^r_{16} + 16/3\,m_\pi^6\,C^r_{17} + 128/9\,m_\pi^6\,C^r_{18} - 16\,
  m_\pi^6\,C^r_{19} + 16\,m_\pi^6\,C^r_{20}
\nonumber\\&&
 - 16\,m_\pi^6\,C^r_{21} - 32/3\,m_\pi^6\,C^r_{31} + 32\,
  m_\pi^6\,C^r_{32} - 128/9\,m_\pi^4\,m_K^2\,C^r_{12} + 64/3\,m_\pi^4\,m_K^2\,C^r_{13}
\nonumber\\&&
 - 320/9\,m_\pi^4
  \,m_K^2\,C^r_{14} 
+ 32/3\,m_\pi^4\,m_K^2\,C^r_{15} - 256/3\,m_\pi^4\,m_K^2\,C^r_{16} 
- 320/9\,m_\pi^4\,
  m_K^2\,C^r_{17}
\nonumber\\&&
 - 256/3\,m_\pi^4\,m_K^2\,C^r_{18} + 192\,m_\pi^4\,m_K^2\,C^r_{19}
 + 64\,m_\pi^4\,m_K^2\,
  C^r_{20} + 128\,m_\pi^4\,m_K^2\,C^r_{31}
\nonumber\\&&
 - 64/3\,m_\pi^4\,m_K^2\,C^r_{32} + 512/3\,m_\pi^4\,m_K^2\,
  C^r_{33} + 512/9\,m_\pi^2\,m_K^4\,C^r_{12} + 640/9\,m_\pi^2\,m_K^4\,C^r_{14}
\nonumber\\&&
 + 320/3\,m_\pi^2\,m_K^4\,
  C^r_{16} + 640/9\,m_\pi^2\,m_K^4\,C^r_{17} + 128\,m_\pi^2\,m_K^4\,C^r_{18} - 384\,m_\pi^2\,m_K^4\,C^r_{19}
\nonumber\\&&
   - 192\,m_\pi^2\,m_K^4\,C^r_{20} + 192\,m_\pi^2\,m_K^4\,C^r_{21} - 256\,m_\pi^2\,m_K^4\,C^r_{31} - 256/
  3\,m_\pi^2\,m_K^4\,C^r_{32}
\nonumber\\&&
 - 1024/3\,m_\pi^2\,m_K^4\,C^r_{33} - 2048/27\,m_K^6\,C^r_{12} - 1024/
  9\,m_K^6\,C^r_{13} - 512/9\,m_K^6\,C^r_{14}
\nonumber\\&&
 - 512/9\,m_K^6\,C^r_{15} - 256/3\,m_K^6\,C^r_{16}
   - 512/9\,m_K^6\,C^r_{17} - 512/9\,m_K^6\,C^r_{18} + 256\,m_K^6\,C^r_{19}
 \nonumber\\&&
+ 256\,m_K^6\,
  C^r_{20} + 256\,m_K^6\,C^r_{21} + 512/3\,m_K^6\,C^r_{31} + 512/3\,m_K^6\,C^r_{32} + 512/3\,
  m_K^6\,C^r_{33} \,,
\nonumber\\&&
\end{eqnarray}
\begin{eqnarray}
\lefteqn{
F_\pi^4 (m_\eta^2)^{(6)}_{Loops} =
\frac{1}{(16\pi^2)^2} \bigg( 
 - 91/11664\,m_\pi^6\,\pi^2 - 1781/15552\,m_\pi^6 - 269/486\,
  m_\pi^4\,m_K^2\,\pi^2 
}&&
\nonumber\\&&
- 4133/1296\,m_\pi^4\,m_K^2 + 133/108\,m_\pi^2\,m_K^4\,\pi^2 + 367/
  48\,m_\pi^2\,m_K^4 - 1091/729\,m_K^6\,\pi^2
\nonumber\\&&
 - 7567/972\,m_K^6 \bigg)
+\frac{1}{16\pi^2}\bigg(\mu_K \, ( 2/3\,m_\pi^4 - 4/3\,m_\pi^2\,m_K^2 - 16/3\,m_K^4 )
+L^r_1 \, (  - 4/27\,m_\pi^6 
\nonumber\\&&
+ 16/9\,m_\pi^4\,m_K^2 - 64/9\,m_\pi^2\,m_K^4 + 
 256/27\,m_K^6 )
+L^r_2 \, (  - 58/27\,m_\pi^6 + 88/9\,m_\pi^4\,m_K^2 
\nonumber\\&&
- 88/9\,m_\pi^2\,m_K^4
 +   544/27\,m_K^6 )
+L^r_3 \, (  - 20/27\,m_\pi^6 + 32/9\,m_\pi^4\,m_K^2 - 34/9\,m_\pi^2\,m_K^4 
\nonumber\\&&
+   152/27\,m_K^6 )\bigg)
+\mu_\pi^2 \, ( 43/12\,m_\pi^2 - 3\,m_K^2 )
+\mu_\pi\,\mu_K \, (  - 8/3\,m_\pi^2 - 32/3\,m_K^2 )
\nonumber\\&&
+\mu_\pi\,\mu_\eta \, (  - 35/9\,m_\pi^2 + 64/9\,m_K^2 )
+\mu_\pi\,L^r_1 \, ( 32\,m_\pi^4 - 128\,m_\pi^2\,m_K^2 )
+\mu_\pi\,L^r_2 \, ( 8\,m_\pi^4
\nonumber\\&&
 - 32\,m_\pi^2\,m_K^2 )
+\mu_\pi\,L^r_3 \, ( 8\,m_\pi^4 - 32\,m_\pi^2\,m_K^2 )
+\mu_\pi\,L^r_4 \, (  - 160/3\,m_\pi^4 + 144\,m_\pi^2\,m_K^2
\nonumber\\&&
 + 256/3\,m_K^4 )
+\mu_\pi\,L^r_5 \, ( 16/9\,m_\pi^4 - 64/3\,m_\pi^2\,m_K^2 + 512/9\,m_K^4 )
\nonumber\\&&
+\mu_\pi\,L^r_6 \, ( 64\,m_\pi^4 - 544/3\,m_\pi^2\,m_K^2 - 512/3\,m_K^4 )
+\mu_\pi\,L^r_7 \, (  - 384\,m_\pi^4 + 1664/3\,m_\pi^2\,m_K^2
\nonumber\\&&
 - 512/3\,m_K^4 )
+\mu_\pi\,L^r_8 \, (  - 144\,m_\pi^4 + 640/3\,m_\pi^2\,m_K^2 - 512/3\,m_K^4 )
\nonumber\\&&
+\mu_K^2 \, (  - 4\,m_\pi^4\,m_K^{-2} + 43/3\,m_\pi^2 - 212/9\,m_K^2 )
+\mu_K\,\mu_\eta \, ( 28/9\,m_\pi^2 - 64/9\,m_K^2 )
\nonumber\\&&
+\mu_K\,L^r_1 \, ( 128/3\,m_\pi^2\,m_K^2 - 512/3\,m_K^4 )
+\mu_K\,L^r_2 \, ( 32/3\,m_\pi^2\,m_K^2 - 128/3\,m_K^4 )
\nonumber\\&&
+\mu_K\,L^r_3 \, ( 56/3\,m_\pi^2\,m_K^2 - 224/3\,m_K^4 )
+\mu_K\,L^r_4 \, (  - 16/3\,m_\pi^4 - 32/3\,m_\pi^2\,m_K^2
\nonumber\\&&
 + 256\,m_K^4 )
+\mu_K\,L^r_5 \, ( 80/9\,m_\pi^4 - 32\,m_\pi^2\,m_K^2 + 1024/9\,m_K^4 )
\nonumber\\&&
+\mu_K\,L^r_6 \, ( 32/3\,m_\pi^4 + 64/3\,m_\pi^2\,m_K^2 - 256\,m_K^4 )
+\mu_K\,L^r_7 \, (  - 512/3\,m_\pi^4 
\nonumber\\&&
+ 1280/3\,m_\pi^2\,m_K^2 
- 256\,m_K^4 )
+\mu_K\,L^r_8 \, (  - 224/3\,m_\pi^4 + 704/3\,m_\pi^2\,m_K^2 - 256\,m_K^4 )
\nonumber\\&&
+\mu_\eta^2 \, (  - 5/9\,m_\pi^4\,m_\eta^{-2} + 55/12\,m_\pi^2 - 25/3\,m_K^2 )
+\mu_\eta\,L^r_1 \, (  - 16/3\,m_\pi^4 + 128/3\,m_\pi^2\,m_K^2
\nonumber\\&&
 - 256/3\,m_K^4 )
+\mu_\eta\,L^r_2 \, (  - 16/3\,m_\pi^4 + 128/3\,m_\pi^2\,m_K^2 - 256/3\,m_K^4 )
\nonumber\\&&
+\mu_\eta\,L^r_3 \, (  - 8/3\,m_\pi^4 + 64/3\,m_\pi^2\,m_K^2 - 128/3\,m_K^4 )
\nonumber\\&&
+\mu_\eta\,L^r_4 \, ( 32/3\,m_\pi^4 - 112/3\,m_\pi^2\,m_K^2 + 128/3\,m_K^4 )
\nonumber\\&&
+\mu_\eta\,L^r_5 \, ( 496/27\,m_\pi^4 - 1472/27\,m_\pi^2\,m_K^2 + 1408/27\,m_K^4 )
\nonumber\\&&
+\mu_\eta\,L^r_6 \, (  - 128/9\,m_\pi^4 + 736/9\,m_\pi^2\,m_K^2 - 2048/9\,m_K^4 )
\nonumber\\&&
+\mu_\eta\,L^r_7 \, (  - 1024/9\,m_\pi^4 + 3584/9\,m_\pi^2\,m_K^2 - 2560/9\,m_K^4 )
\nonumber\\&&
+\mu_\eta\,L^r_8 \, (  - 592/9\,m_\pi^4 + 2176/9\,m_\pi^2\,m_K^2 - 256\,m_K^4 )
\nonumber\\&&
+L^{r2}_4 \, ( 128/3\,m_\pi^6 - 512\,m_\pi^2\,m_K^4 - 2048/3\,m_K^6 )
\nonumber\\&&
+L^r_4\,L^r_5 \, ( 256/9\,m_\pi^6 - 128/3\,m_\pi^4\,m_K^2 - 128\,m_\pi^2\,m_K^4 - 5632/
  9\,m_K^6 )
\nonumber\\&&
+L^r_4\,L^r_6 \, (  - 512/3\,m_\pi^6 + 2048\,m_\pi^2\,m_K^4 + 8192/3\,m_K^6 )
\nonumber\\&&
+L^r_4\,L^r_7 \, ( 1024\,m_\pi^6 - 3072\,m_\pi^2\,m_K^4 + 2048\,m_K^6 )
\nonumber\\&&
+L^r_4\,L^r_8 \, ( 1024/3\,m_\pi^6 - 256/3\,m_\pi^4\,m_K^2 - 3328/3\,m_\pi^2\,m_K^4 + 
  7168/3\,m_K^6 )
\nonumber\\&&
+L^{r2}_5 \, (  - 128/9\,m_\pi^6 + 256/3\,m_\pi^4\,m_K^2 - 256/3\,m_\pi^2\,m_K^4 - 
  1024/9\,m_K^6 )
\nonumber\\&&
+L^r_5\,L^r_6 \, (  - 1024/9\,m_\pi^6 + 256/3\,m_\pi^4\,m_K^2 + 1280/3\,m_\pi^2\,m_K^4
   + 10240/9\,m_K^6 )
\nonumber\\&&
+L^r_5\,L^r_7 \, ( 1024\,m_\pi^6 - 4096/3\,m_\pi^4\,m_K^2 - 1024/3\,m_\pi^2\,m_K^4 + 
  2048/3\,m_K^6 )
\nonumber\\&&
+L^r_5\,L^r_8 \, ( 3584/9\,m_\pi^6 - 6656/9\,m_\pi^4\,m_K^2 - 512/9\,m_\pi^2\,m_K^4 + 
  8192/9\,m_K^6 )
\nonumber\\&&
+L^{r2}_6 \, ( 512/3\,m_\pi^6 - 2048\,m_\pi^2\,m_K^4 - 8192/3\,m_K^6 )
\nonumber\\&&
+L^r_6\,L^r_7 \, (  - 4096/3\,m_\pi^6 + 4096\,m_\pi^2\,m_K^4 - 8192/3\,m_K^6 )
\nonumber\\&&
+L^r_6\,L^r_8 \, (  - 1024/3\,m_\pi^6 + 512/3\,m_\pi^4\,m_K^2 + 3584/3\,m_\pi^2\,m_K^4
   - 4096\,m_K^6 )
\nonumber\\&&
+L^r_7\,L^r_8 \, (  - 4096/3\,m_\pi^6 + 4096/3\,m_\pi^4\,m_K^2 + 4096/3\,m_\pi^2\,
  m_K^4 - 4096/3\,m_K^6 )
\nonumber\\&&
+L^{r2}_8 \, (  - 512\,m_\pi^6 + 2048/3\,m_\pi^4\,m_K^2 + 2048/3\,m_\pi^2\,m_K^4 - 
  4096/3\,m_K^6 )
\nonumber\\&&
+  1/6\,H^F(m_\pi^2,m_\pi^2,m_\eta^2;m_\eta^2)\,m_\pi^4 + 1/8\,H^F(m_\pi^2,m_K^2,m_K^2;m_\eta^2)\,
  m_\pi^4
\nonumber\\&&
 + 3/2\,H^F(m_\pi^2,m_K^2,m_K^2;m_\eta^2)\,m_\pi^2\,m_K^2 - 1/2\,H^F(m_K^2,m_K^2,m_\eta^2;m_\eta^2)\,
  m_\pi^4
\nonumber\\&&
 + 4\,H^F(m_K^2,m_K^2,m_\eta^2;m_\eta^2)\,m_\pi^2\,m_K^2 - 64/9\,H^F(m_K^2,m_K^2,m_\eta^2;m_\eta^2)\,
  m_K^4
\nonumber\\&&
 + 1/8\,H^F(m_\eta^2,m_K^2,m_K^2;m_\eta^2)\,m_\pi^4 - H^F(m_\eta^2,m_K^2,m_K^2;m_\eta^2)\,m_\pi^2\,m_K^2
\nonumber\\&&
   + 2\,H^F(m_\eta^2,m_K^2,m_K^2;m_\eta^2)\,m_K^4
 + 49/486\,H^F(m_\eta^2,m_\eta^2,m_\eta^2;m_\eta^2)\,m_\pi^4
\nonumber\\&&
   - 112/243\,H^F(m_\eta^2,m_\eta^2,m_\eta^2;m_\eta^2)\,m_\pi^2\,m_K^2 + 128/243\,H^F(m_\eta^2,m_\eta^2,m_\eta^2;m_\eta^2
  )\,m_K^4
\nonumber\\&&
 + H^F_1(m_\pi^2,m_K^2,m_K^2;m_\eta^2)\,m_\pi^4 - 4\,H^F_1(m_\pi^2,m_K^2,m_K^2;m_\eta^2)\,m_\pi^2\,
  m_K^2 
\nonumber\\&&
+ 2\,H^F_1(m_K^2,m_K^2,m_\eta^2;m_\eta^2)\,m_\pi^4 - 40/3\,H^F_1(m_K^2,m_K^2,m_\eta^2;m_\eta^2)\,m_\pi^2\,
  m_K^2
\nonumber\\&&
 + 64/3\,H^F_1(m_K^2,m_K^2,m_\eta^2;m_\eta^2)\,m_K^4 + 3/8\,H^F_{21}(m_\pi^2,m_K^2,m_K^2;m_\eta^2)\,
  m_\pi^4
\nonumber\\&&
 - 3\,H^F_{21}(m_\pi^2,m_K^2,m_K^2;m_\eta^2)\,m_\pi^2\,m_K^2 + 6\,H^F_{21}(m_\pi^2,m_K^2,m_K^2;m_\eta^2)\,
  m_K^4
\nonumber\\&&
 + 3/8\,H^F_{21}(m_\eta^2,m_K^2,m_K^2;m_\eta^2)\,m_\pi^4 - 3\,H^F_{21}(m_\eta^2,m_K^2,m_K^2;m_\eta^2)\,
  m_\pi^2\,m_K^2 
\nonumber\\&&
+ 6\,H^F_{21}(m_\eta^2,m_K^2,m_K^2;m_\eta^2)\,m_K^4 )\,.
\end{eqnarray}

\subsection{Decay Constants}
\label{appdecay}

The decay constants are given by
\begin{equation}
F_a = F_0 \left(1 +\bar F_a^{(4)} +(\bar F_a^{(6)})_{CT} 
 +(\bar F_a^{(6)})_{loops} \right)\,.
\end{equation}

For the pion we obtain
\begin{equation}
F_\pi^2 \bar F_\pi^{(4)} = 4 (m_\pi^2+2m_K^2)L^r_4 + 4 m_\pi^2 L^r_5
 -2 \mu_\pi - \mu_K
\end{equation}
in agreement with \cite{GL} for the \pfour\ contribution.
The \psix\ contribution is
\begin{eqnarray}
\lefteqn{
F_\pi^2(\bar F_\pi)^{(6)}_{CT} =
  8\,m_\pi^4\,C^r_{14} + 8\,m_\pi^4\,C^r_{15} + 24\,m_\pi^4\,C^r_{16} + 8\,m_\pi^4\,C^r_{17}
}&&
\nonumber\\&&
    + 16\,m_\pi^2\,m_K^2\,C^r_{15} - 32\,m_\pi^2\,m_K^2\,C^r_{16} + 32\,m_K^4\,C^r_{16} \,,
\end{eqnarray}
\begin{eqnarray}
\lefteqn{
F_\pi^4(\bar F_\pi^{(6)})_{loops} =  \frac{1}{(16\pi^2)^2} \, 
\bigg( 35/288\,m_\pi^4\,\pi^2 + 41/128\,m_\pi^4 + 1/144\,m_\pi^2\,m_K^2\,\pi^2
}&&
\nonumber\\&&
    - 5/32\,m_\pi^2\,m_K^2 + 11/72\,m_K^4\,\pi^2 + 15/32\,m_K^4 \bigg)
 + \frac{1}{16\pi^2}\,\bigg(\mu_\pi \, ( 1/2\,m_\pi^2 + m_K^2 )
\nonumber\\&&
 +( 1/4\,m_\pi^2 + m_K^2 ) \, \mu_K 
 +( 1/3\,m_\pi^2 - 1/3\,m_K^2 ) \, \mu_\eta 
 - 2\,m_\pi^4 \, L_1^r 
 + (  - 37/9\,m_\pi^4 
\nonumber\\&&
+ 8/9\,m_\pi^2\,m_K^2 - 52/9\,m_K^4 )\, L_2^r 
 + (  - 28/27\,m_\pi^4 + 8/27\,m_\pi^2\,m_K^2 - 43/27\,m_K^4 ) \, L_3^r
\nonumber\\&&
 +( 8\,m_\pi^4 + 20\,m_\pi^2\,m_K^2 + 8\,m_K^4 )\,  L_4^r 
 + ( 8\,m_\pi^4 + 4\,m_K^4 ) \, L_5^r
 + (  - 16\,m_\pi^4
\nonumber\\&&
 - 40\,m_\pi^2\,m_K^2 - 16\,m_K^4 ) \, L_6^r
 +(  - 16\,m_\pi^4 - 8\,m_K^4 ) \,  L_8^r \, \bigg)
 +( 7/8 + m_\pi^{-2}\,m_K^2 ) \, \mu_\pi^2 
\nonumber\\&&
 + 2/3 \, \mu_\pi\,\mu_\eta 
 +  56\,m_\pi^2 \, \mu_\pi\,L_1^r 
 + 32\,m_\pi^2  \,\mu_\pi\,L_2^r 
 +  28\,m_\pi^2  \, \mu_\pi\,L_3^r 
\nonumber\\&&
 + 48\,m_K^2 \,\mu_\pi\,L_4^r 
 + 12\,m_\pi^2 \,  \mu_\pi\,L_5^r 
 +(  - 32\,m_\pi^2 - 64\,m_K^2 ) \,  \mu_\pi\,L_6^r
- 32\,m_\pi^2 \,\mu_\pi\,L_8^r 
\nonumber\\&&
 +( 3 - 1/2\,m_\pi^2\,m_K^{-2} ) \,  \mu_K^2
 - 2/3 \, \mu_K\,\mu_\eta 
 +64\,m_K^2  \, \mu_K\,L_1^r 
 + 16\,m_K^2 \, \mu_K\,L_2^r 
\nonumber\\&&
 + 20\,m_K^2 \, \mu_K\,L_3^r 
 +( 12\,m_\pi^2 - 8\,m_K^2 ) \,  \mu_K\,L_4^r
 +(  - 4\,m_\pi^2 + 8\,m_K^2 )  \, \mu_K\,L_5^r
\nonumber\\&&
 +(  - 16\,m_\pi^2 - 32\,m_K^2 )  \, \mu_K\,L_6^r
 - 16\,m_K^2  \, \mu_K\,L_8^r 
 + 3/8 \, \mu_\eta^2 
\nonumber\\&&
 + (  - 16/3\,m_\pi^2 + 64/3\,m_K^2 )  \,\mu_\eta\,L_1^r
 + (  - 4/3\,m_\pi^2 + 16/3\,m_K^2 ) \, \mu_\eta\,L_2^r
\nonumber\\&&
 +(  - 4/3\,m_\pi^2 + 16/3\,m_K^2 ) \,  \mu_\eta\,L_3^r
 +( 8/3\,m_\pi^2 - 32/3\,m_K^2 ) \,  \mu_\eta\,L_4^r
- 4/3\,m_\pi^2 \, \mu_\eta\,L_5^r 
\nonumber\\&&
 +(  - 8\,m_\pi^4 - 32\,m_\pi^2\,m_K^2 - 32\,m_K^4 ) \,  L_4^{r2}
 +(  - 16\,m_\pi^4 - 32\,m_\pi^2\,m_K^2 ) \,  L_4^r\,L_5^r
\nonumber\\&&
 - 8\,m_\pi^4 \, L_5^{r2} 
 + 5/12\,H^{F\prime}(m_\pi^2,m_\pi^2,m_\pi^2;m_\pi^2)\,m_\pi^4 
 - 1/2\,H^F(m_\pi^2,m_\pi^2,m_\pi^2;m_\pi^2)\,m_\pi^2 
\nonumber\\&&
- 5/16\,H^{F\prime}(m_\pi^2,m_K^2,m_K^2;m_\pi^2)\,m_\pi^4
 + 1/16\,H^F(m_\pi^2,m_K^2,m_K^2;m_\pi^2)\,m_\pi^2
\nonumber\\&&
 + 1/36\,H^{F\prime}(m_\pi^2,m_\eta^2,m_\eta^2;m_\pi^2)\,m_\pi^4 
 + 1/2\,H^{F\prime}(m_K^2,m_\pi^2,m_K^2;m_\pi^2)\,m_\pi^2\,m_K^2 
\nonumber\\&&
 - 1/2\,H^F(m_K^2,m_\pi^2,m_K^2;m_\pi^2)\,m_K^2
 - 5/12\,H^{F\prime}(m_K^2,m_K^2,m_\eta^2;m_\pi^2)\,m_\pi^4 
\nonumber\\&&
- 1/16\,H^{F\prime}(m_\eta^2,m_K^2,m_K^2;m_\pi^2)\,m_\pi^4
 + 1/4\,H^{F\prime}(m_\eta^2,m_K^2,m_K^2;m_\pi^2)\,m_\pi^2\,m_K^2 
\nonumber\\&&
+ 1/16\,H^F(m_\eta^2,m_K^2,m_K^2;m_\pi^2)\,m_\pi^2
 - 1/4\,H^F(m_\eta^2,m_K^2,m_K^2;m_\pi^2)\,m_K^2 
\nonumber\\&&
+ 1/2\,H^{F\prime}_1(m_\pi^2,m_K^2,m_K^2;m_\pi^2)\,m_\pi^4
 + H^{F\prime}_1(m_K^2,m_K^2,m_\eta^2;m_\pi^2)\,m_\pi^4 
\nonumber\\&&
+ 3/2\,H^{F\prime}_{21}(m_\pi^2,m_\pi^2,m_\pi^2;m_\pi^2)\,m_\pi^4
 - 3/16\,H^{F\prime}_{21}(m_\pi^2,m_K^2,m_K^2;m_\pi^2)\,m_\pi^4 
\nonumber\\&&
+ 3/2\,H^{F\prime}_{21}(m_K^2,m_\pi^2,m_K^2;m_\pi^2)\,m_\pi^4
 + 9/16\,H^{F\prime}_{21}(m_\eta^2,m_K^2,m_K^2;m_\pi^2)\,m_\pi^4 \,.
\end{eqnarray}
\noindent
The functions $H^{F\prime}_i(m_1^2,m_2^2,m_3^2;q^2)$ are defined in 
App. \ref{loopint}.

For the kaon we obtain
\begin{equation}
F_\pi^2 \bar F_K^{(4)} = 4 (m_\pi^2+2m_K^2)L^r_4 + 4 m_K^2 L^r_5
 -\frac{3}{4} \mu_\pi - \frac{3}{2}\mu_K -\frac{3}{4}\mu_\eta\,.
\end{equation}
The \psix\ contribution is given by
\begin{eqnarray}
\lefteqn{
F_\pi^2(\bar F_K^{(6)})_{CT} = 
  8\,m_\pi^4\,C^r_{14} + 24\,m_\pi^4\,C^r_{16} - 8\,m_\pi^4\,C^r_{17}
 - 16\,m_\pi^2\,m_K^2\,
   C^r_{14}
 + 8\,m_\pi^2\,m_K^2\,C^r_{15} 
}&&
\nonumber\\&&
- 32\,m_\pi^2\,m_K^2\,C^r_{16} + 16\,m_\pi^2\,m_K^2\,C^r_{17} + 16\,m_K^4\,
   C^r_{14} + 16\,m_K^4\,C^r_{15} + 32\,m_K^4\,C^r_{16} \,,\hskip1cm
\end{eqnarray}
\begin{eqnarray}
\lefteqn{
F_\pi^4(\bar F_K^{(6)})_{Loops} =
  \frac{1}{(16\pi^2)} \, \bigg( 1/16\,m_\pi^4\,\pi^2 + 13/48\,m_\pi^4 + 1/32\,m_\pi^2\,m_K^2\,\pi^2 
}&&
\nonumber\\&&
-    29/192\,m_\pi^2\,m_K^2 + 3/16\,m_K^4\,\pi^2 + 197/384\,m_K^4 \bigg)
 + \frac{1}{16\pi^2}\bigg( 15/16\,m_K^2  \, \mu_\pi 
\nonumber\\&&
 + ( 3/4\,m_\pi^2 + 1/8\,m_K^2 ) \, \mu_K
 +( 1/12\,m_\pi^2 + 41/48\,m_K^2 ) \,  \mu_\eta
- 2\,m_K^4 \,   L_1^r
\nonumber\\&&
 +(  - 28/9\,m_\pi^4 + 8/9\,m_\pi^2\,m_K^2 - 61/9\,m_K^4 ) \, L_2^r 
 + (  - 41/54\,m_\pi^4 + 2/27\,m_\pi^2\,m_K^2 
\nonumber\\&&
- 89/54\,m_K^4 ) \, L_3^r
 +( 2\,m_\pi^4 + 14\,m_\pi^2\,m_K^2 + 20\,m_K^4 ) \,  L_4^r
 +( 10/3\,m_\pi^4 - 8/3\,m_\pi^2\,m_K^2 
\nonumber\\&&
+ 34/3\,m_K^4 ) \,  L_5^r
 +(  - 4\,m_\pi^4 - 28\,m_\pi^2\,m_K^2 - 40\,m_K^4 ) \,  L_6^r
 +(  - 16\,m_\pi^4 + 32\,m_\pi^2\,m_K^2 
\nonumber\\&&
- 16\,m_K^4 )  \, L_7^r
 +(  - 12\,m_\pi^4 + 16\,m_\pi^2\,m_K^2 - 28\,m_K^4 ) \,  L_8^r\bigg)
 + \mu_\pi^2 \, ( 9/32 + 3/8\,m_\pi^{-2}\,m_K^2 ) 
\nonumber\\&&
 +  3/8  \, \mu_\pi\,\mu_K
 +( 13/16 + 3/4\,m_\pi^2\,m_\eta^{-2} ) \,  \mu_\pi\,\mu_\eta
 + 48\,m_\pi^2 \, \mu_\pi\,L_1^r
 + 12\,m_\pi^2 \, \mu_\pi\,L_2^r 
\nonumber\\&&
 + 15\,m_\pi^2  \, \mu_\pi\,L_3^r
 +(  - 15\,m_\pi^2 + 18\,m_K^2 ) \,  \mu_\pi\,L_4^r
 + ( 6\,m_\pi^2 - 3\,m_K^2 )  \, \mu_\pi\,L_5^r
\nonumber\\&&
 +(  - 12\,m_\pi^2 - 24\,m_K^2 ) \,  \mu_\pi\,L_6^r
 - 12\,m_\pi^2 \, \mu_\pi\,L_8^r
 + \mu_K^2 \, ( 27/8 + 3/4\,m_\pi^2\,m_K^{-2} )
\nonumber\\&&
 +(  - 29/8 - 1/2\,m_\pi^2\,m_\eta^{-2} )  \,\mu_K\,\mu_\eta 
 +72\,m_K^2  \,  \mu_K\,L_1^r
 + 36\,m_K^2 \,\mu_K\,L_2^r 
\nonumber\\&&
 + 30\,m_K^2 \, \mu_K\,L_3^r 
 +( 18\,m_\pi^2 + 4\,m_K^2 )  \, \mu_K\,L_4^r
 + 6\,m_K^2 \, \mu_K\,L_5^r 
\nonumber\\&&
 + (  - 24\,m_\pi^2 - 48\,m_K^2 ) \, \mu_K\,L_6^r
- 24\,m_K^2 \, \mu_K\,L_8^r 
 +( 17/8 + 1/32\,m_\pi^2\,m_\eta^{-2} ) \,  \mu_\eta^2
\nonumber\\&&
 +(  - 16/3\,m_\pi^2 + 64/3\,m_K^2 ) \,  \mu_\eta\,L_1^r
 +(  - 4/3\,m_\pi^2 + 16/3\,m_K^2 ) \,  \mu_\eta\,L_2^r
\nonumber\\&&
 +(  - 7/3\,m_\pi^2 + 28/3\,m_K^2 ) \,  \mu_\eta\,L_3^r
 +( 35/3\,m_\pi^2 + 22/3\,m_K^2 ) \,  \mu_\eta\,L_4^r
\nonumber\\&&
 + (  - 2\,m_\pi^2 + 23/3\,m_K^2 ) \, \mu_\eta\,L_5^r
 +(  - 12\,m_\pi^2 - 24\,m_K^2 ) \,  \mu_\eta\,L_6^r
 +(  - 18\,m_\pi^4\,m_\eta^{-2} 
\nonumber\\&&
+ 42\,m_\pi^2 - 24\,m_K^2 ) \,  \mu_\eta\,L_7^r
 +(  - 6\,m_\pi^4\,m_\eta^{-2} + 18\,m_\pi^2 - 24\,m_K^2 ) \,  \mu_\eta\,L_8^r
 +(  - 8\,m_\pi^4 
\nonumber\\&&
- 32\,m_\pi^2\,m_K^2 - 32\,m_K^4 ) \,  L_4^{r2}
 +(  - 16\,m_\pi^2\,m_K^2 - 32\,m_K^4 ) \,  L_4^r\,L_5^r
- 8\,m_K^4 \,  L_5^{r2} 
\nonumber\\&&
 +  3/8\,H^{F\prime}(m_\pi^2,m_\pi^2,m_K^2;m_K^2)\,m_\pi^2\,m_K^2
 + 3/16\,H^{F\prime}(m_\pi^2,m_\pi^2,m_K^2;m_K^2)\,m_K^4
\nonumber\\&&
 - 3/8\,H^{F\prime}(m_\pi^2,m_\pi^2,m_K^2;m_K^2)\,m_\pi^2
 + 1/8\,H^{F\prime}(m_\pi^2,m_K^2,m_\eta^2;m_K^2)\,m_K^4
\nonumber\\&&
 - 3/64\,H^{F\prime}(m_K^2,m_\pi^2,m_\pi^2;m_K^2)\,m_K^4
 + 3/64\,H^F(m_K^2,m_\pi^2,m_\pi^2;m_K^2)\,m_K^2
\nonumber\\&&
 + 9/32\,H^{F\prime}(m_K^2,m_\pi^2,m_\eta^2;m_K^2)\,m_K^4
 - 9/32\,H^F(m_K^2,m_\pi^2,m_\eta^2;m_K^2)\,m_K^2
\nonumber\\&&
 + 3/8\,H^{F\prime}(m_K^2,m_K^2,m_K^2;m_K^2)\,m_K^4
 - 3/8\,H^F(m_K^2,m_K^2,m_K^2;m_K^2)\,m_K^2
\nonumber\\&&
 + 181/576\,H^{F\prime}(m_K^2,m_\eta^2,m_\eta^2;m_K^2)\,m_K^4
 - 9/64\,H^F(m_K^2,m_\eta^2,m_\eta^2;m_K^2)\,m_K^2 
\nonumber\\&&
- 3/4\,H^{F\prime}_1(m_\pi^2,m_\pi^2,m_K^2;m_K^2)\,m_K^4
 - 3/4\,H^{F\prime}_1(m_K^2,m_\pi^2,m_\eta^2;m_K^2)\,m_K^4 
\nonumber\\&&
- 5/8\,H^{F\prime}_1(m_K^2,m_\eta^2,m_\eta^2;m_K^2)\,m_K^4
 + 9/8\,H^{F\prime}_{21}(m_\pi^2,m_\pi^2,m_K^2;m_K^2)\,m_K^4
\nonumber\\&&
 - 9/64\,H^{F\prime}_{21}(m_K^2,m_\pi^2,m_\pi^2;m_K^2)\,m_K^4
 + 27/32\,H^{F\prime}_{21}(m_K^2,m_\pi^2,m_\eta^2;m_K^2)\,m_K^4 
\nonumber\\&&
+ 9/8\,H^{F\prime}_{21}(m_K^2,m_K^2,m_K^2;m_K^2)\,m_K^4
 + 27/64\,H^{F\prime}_{21}(m_K^2,m_\eta^2,m_\eta^2;m_K^2)\,m_K^4 \,.
\end{eqnarray}

For the eta we obtain
\begin{equation}
F_\pi^2 \bar F_\eta^{(4)} = 4 (m_\pi^2+2m_K^2)L^r_4 
+ \frac{4}{3} (4m_K^2-m_\pi^2) L^r_5- 3\mu_K
\end{equation}
in agreement with \cite{GL} for the \pfour\ contribution.
The \psix\ contribution is
\begin{eqnarray}
\lefteqn{
F_\pi^2(\bar F_\eta^{(6)})_{CT} = 
  8\,m_\pi^4\,C^r_{14} - 8/3\,m_\pi^4\,C^r_{15} + 24\,m_\pi^4\,C^r_{16} + 8\,m_\pi^4\,C^r_{17}
    + 64/3\,m_\pi^4\,C^r_{18}
}&&
\nonumber\\&&
 - 64/3\,m_\pi^2\,m_K^2\,C^r_{14} + 16/3\,m_\pi^2\,m_K^2\,C^r_{15} - 32\,m_\pi^2\,
   m_K^2\,C^r_{16} - 64/3\,m_\pi^2\,m_K^2\,C^r_{17}
\nonumber\\&&
 - 128/3\,m_\pi^2\,m_K^2\,C^r_{18} + 64/3\,m_K^4\,C^r_{14}
    + 64/3\,m_K^4\,C^r_{15} + 32\,m_K^4\,C^r_{16} 
\nonumber\\&&
+ 64/3\,m_K^4\,C^r_{17} + 64/3\,m_K^4\,
   C^r_{18} \,,
\end{eqnarray}
\begin{eqnarray}
\lefteqn{
F_\pi^4(\bar F_\eta^{(6)})_{Loops} =
  \frac{1}{(16\pi^2)} \, \bigg( 5/96\,m_\pi^4\,\pi^2 + 91/384\,m_\pi^4 + 1/48\,m_\pi^2\,m_K^2\,\pi^2 
}&&
\nonumber\\&&
-   11/96\,m_\pi^2\,m_K^2 + 5/24\,m_K^4\,\pi^2 + 49/96\,m_K^4 \bigg)
 + \frac{1}{16\pi^2}\bigg( ( 3/4\,m_\pi^2 + 3\,m_K^2 ) \, \mu_K 
\nonumber\\&&
 - m_K^2\, \mu_\eta 
 +(  - 2/9\,m_\pi^4 + 16/9\,m_\pi^2\,m_K^2 - 32/9\,m_K^4 ) \,  L_1^r
 +(  - 29/9\,m_\pi^4 
\nonumber\\&&
 + 16/9\,m_\pi^2\,m_K^2 - 68/9\,m_K^4 ) \,  L_2^r
 +(  - 10/9\,m_\pi^4 + 8/9\,m_\pi^2\,m_K^2 - 19/9\,m_K^4 ) \,  L_3^r
\nonumber\\&&
 + ( 12\,m_\pi^2\,m_K^2 + 24\,m_K^4 ) \, L_4^r
 +12\,m_K^4  \, L_5^r
 +(  - 24\,m_\pi^2\,m_K^2 - 48\,m_K^4 ) \,  L_6^r
\nonumber\\&&
 - 24\,m_K^4 \,  L_8^r\bigg)
 + 9/8  \, \mu_\pi^2
 + 48\,m_\pi^2 \, \mu_\pi\,L_1^r
 + 12\,m_\pi^2 \,\mu_\pi\,L_2^r 
 + 12\,m_\pi^2 \, \mu_\pi\,L_3^r 
\nonumber\\&&
 - 24\,m_\pi^2 \, \mu_\pi\,L_4^r 
 - 4\,m_\pi^2 \,\mu_\pi\,L_5^r 
 + ( 3 + 3/2\,m_\pi^2\,m_K^{-2} )  \, \mu_K^2
 -2 \, \mu_K\,\mu_\eta 
\nonumber\\&&
 + 64\,m_K^2 \, \mu_K\,L_1^r 
 + 16\,m_K^2 \, \mu_K\,L_2^r 
 + 28\,m_K^2 \, \mu_K\,L_3^r 
 + ( 36\,m_\pi^2 + 40\,m_K^2 ) \, \mu_K\,L_4^r
\nonumber\\&&
 + ( 4\,m_\pi^2 + 56/3\,m_K^2 ) \, \mu_K\,L_5^r
 + (  - 48\,m_\pi^2 - 96\,m_K^2 ) \, \mu_K\,L_6^r
- 48\,m_K^2 \, \mu_K\,L_8^r
\nonumber\\&&
 + 9/8 \, \mu_\eta^2 
 + (  - 8\,m_\pi^2 + 32\,m_K^2 ) \, \mu_\eta\,L_1^r
 + (  - 8\,m_\pi^2 + 32\,m_K^2 ) \, \mu_\eta\,L_2^r
\nonumber\\&&
 +(  - 4\,m_\pi^2 + 16\,m_K^2 ) \,  \mu_\eta\,L_3^r
 +( 8/3\,m_\pi^2 - 32/3\,m_K^2 ) \, \mu_\eta\,L_4^r 
 +( 28/9\,m_\pi^2
\nonumber\\&&
 - 64/9\,m_K^2 ) \,  \mu_\eta\,L_5^r
 +(  - 8\,m_\pi^4 - 32\,m_\pi^2\,m_K^2 - 32\,m_K^4 )  \, L_4^{r2}
 +( 16/3\,m_\pi^4 
\nonumber\\&&
  - 32/3\,m_\pi^2\,m_K^2 - 128/3\,m_K^4 ) \,  L_4^r\,L_5^r
 + (  - 8/9\,m_\pi^4 + 64/9\,m_\pi^2\,m_K^2 - 128/9\,m_K^4 )  \,L_5^{r2}
\nonumber\\&&
 +  1/12\,H^{F\prime}(m_\pi^2,m_\pi^2,m_\eta^2;m_\eta^2)\,m_\pi^4
 + 1/16\,H^{F\prime}(m_\pi^2,m_K^2,m_K^2;m_\eta^2)\,m_\pi^4
\nonumber\\&&
 + 3/4\,H^{F\prime}(m_\pi^2,m_K^2,m_K^2;m_\eta^2)\,m_\pi^2\,m_K^2
 - 9/16\,H^F(m_\pi^2,m_K^2,m_K^2;m_\eta^2)\,m_\pi^2
\nonumber\\&&
 - 1/4\,H^{F\prime}(m_K^2,m_K^2,m_\eta^2;m_\eta^2)\,m_\pi^4
 + 2\,H^{F\prime}(m_K^2,m_K^2,m_\eta^2;m_\eta^2)\,m_\pi^2\,m_K^2 
\nonumber\\&&
- 32/9\,H^{F\prime}(m_K^2,m_K^2,m_\eta^2;m_\eta^2)\,m_K^4
 + 1/16\,H^{F\prime}(m_\eta^2,m_K^2,m_K^2;m_\eta^2)\,m_\pi^4 
\nonumber\\&&
- 1/2\,H^{F\prime}(m_\eta^2,m_K^2,m_K^2;m_\eta^2)\,m_\pi^2\,m_K^2
 + H^{F\prime}(m_\eta^2,m_K^2,m_K^2;m_\eta^2)\,m_K^4
\nonumber\\&&
 + 3/16\,H^F(m_\eta^2,m_K^2,m_K^2;m_\eta^2)\,m_\pi^2
 - 3/4\,H^F(m_\eta^2,m_K^2,m_K^2;m_\eta^2)\,m_K^2
\nonumber\\&&
 + 49/972\,H^{F\prime}(m_\eta^2,m_\eta^2,m_\eta^2;m_\eta^2)\,m_\pi^4
 - 56/243\,H^{F\prime}(m_\eta^2,m_\eta^2,m_\eta^2;m_\eta^2)\,m_\pi^2\,m_K^2 
\nonumber\\&&
+ 64/243\,H^{F\prime}(m_\eta^2,m_\eta^2,m_\eta^2;m_\eta^2)\,m_K^4
 + 1/2\,H^{F\prime}_1(m_\pi^2,m_K^2,m_K^2;m_\eta^2)\,m_\pi^4
\nonumber\\&&
 - 2\,H^{F\prime}_1(m_\pi^2,m_K^2,m_K^2;m_\eta^2)\,m_\pi^2\,m_K^2
 +   H^{F\prime}_1(m_K^2,m_K^2,m_\eta^2;m_\eta^2)\,m_\pi^4 
\nonumber\\&&
 - 20/3\,H^{F\prime}_1(m_K^2,m_K^2,m_\eta^2;m_\eta^2)\,m_\pi^2\,m_K^2
    + 32/3\,H^{F\prime}_1(m_K^2,m_K^2,m_\eta^2;m_\eta^2)\,m_K^4
\nonumber\\&&
 + 3/16\,H^{F\prime}_{21}(m_\pi^2,m_K^2,m_K^2;m_\eta^2)\,m_\pi^4
 - 3/2\,H^{F\prime}_{21}(m_\pi^2,m_K^2,m_K^2;m_\eta^2)\,m_\pi^2\,m_K^2 
\nonumber\\&&
+ 3\,H^{F\prime}_{21}(m_\pi^2,m_K^2,m_K^2;m_\eta^2)\,m_K^4
 + 3/16\,H^{F\prime}_{21}(m_\eta^2,m_K^2,m_K^2;m_\eta^2)\,m_\pi^4
\nonumber\\&&
 - 3/2\,H^{F\prime}_{21}(m_\eta^2,m_K^2,m_K^2;m_\eta^2)\,m_\pi^2\,m_K^2
 + 3\,H^{F\prime}_{21}(m_\eta^2,m_K^2,m_K^2;m_\eta^2)\,m_K^4 \,.
\end{eqnarray}

\section{Axial-Vector Two-Point Functions}
\label{axialvector}
\subsection{The remainder transverse part}
\label{transverse}

The remaining transverse pieces are given by
\begin{equation}
\left.\hat\Pi_{Aa}^{(1)}(q^2) =
\left(\hat\Pi_{Aa}^{(1)}(q^2)\right)^{(4)}
+\left(\hat\Pi_{Aa}^{(1)}(q^2)\right)^{(6)}_{CT}
+\left(\hat\Pi_{Aa}^{(1)}(q^2)\right)^{(6)}_{loops}
\right|_{a = \pi,K,\eta} \,.
\end{equation}

The $\pfour$ contribution is the same for all three two-point functions
\begin{equation}
\left(\hat\Pi_{A\pi}^{(1)}(q^2)\right)^{(4)} =
\left(\hat\Pi_{AK}^{(1)}(q^2)\right)^{(4)} =
\left(\hat\Pi_{A\eta}^{(1)}(q^2)\right)^{(4)} =
       4\,L^r_{10} - 8\,H^r_1\,.
\end{equation}

The $\psix$ contribution can be written in a somewhat simpler
fashion by using the function
\begin{eqnarray}
H^T(m_1^2,m_2^2,m_3^2;q^2;m_4^2)
= 3 m_4^4 H^{F\prime}_{21}(m_1^2,m_2^2,m_3^2;m_4^2)
-q^2 H^{F}_{21}(m_1^2,m_2^2,m_3^2;q^2)\,.
\end{eqnarray}

The isospin remainder is
\begin{eqnarray}
\lefteqn{
\left(q^2 \hat\Pi_{A\pi}^{(1)}(q^2)\right)^{(6)}_{CT} =
       - 8\,m_\pi^4\,C^r_{91} - 32\,m_\pi^2\,C^r_{12}\,q^2 
- 32\,m_\pi^2\,C^r_{13}\,q^2 - 32\,m_\pi^2\,C^r_{80}\,q^2 
}&&\nonumber\\&&
- 32\,m_\pi^2\,C^r_{81}\,q^2 - 64\,m_K^2\,C^r_{13}\,q^2 
- 64\,m_K^2\,C^r_{81}\,q^2 - 16\,C^r_{87}\,q^4 - 8\,C^r_{93}\,q^4 \,,
\end{eqnarray}
\begin{eqnarray}
\lefteqn{
\left(q^2 F_\pi^2 \hat\Pi_{A\pi}^{(1)}\right)^{(6)}_{loops} =
 \frac{1}{(16\pi^2)^2} \, 
\bigg( 3/32\,m_\pi^4 + 1/6\,m_\pi^2\,q^2\,\pi^2 
+ 1/3\,m_\pi^2\,q^2
}&&\nonumber\\&&
+ 1/12\,m_K^2\,q^2\,\pi^2 
- 5/24\,m_K^2\,q^2 + 3/32\,q^4 \bigg)
+ \frac{1}{16\pi^2}\,\left( 2\,q^2 \mu_\pi \, 
+ \, q^2 \mu_K \, \right)
\nonumber\\&&
+ 4\,m_\pi^{-2}\,q^2 \mu_\pi^2 \, 
 - 32\,L^r_{10}\,q^2\mu_\pi \, 
+ 2\,m_K^{-2}\,q^2\mu_K^2 \, 
- 16\,L^r_{10}\,q^2\mu_K \, 
\nonumber\\&&
+ 5/3\,m_\pi^4 \, H^{F\prime}(m_\pi^2,m_\pi^2,m_\pi^2;m_\pi^2) \, 
- 2\,m_\pi^2 \,H^F(m_\pi^2,m_\pi^2,m_\pi^2;m_\pi^2) \, 
\nonumber\\&&
+ 2\,m_\pi^2 \,H^F(m_\pi^2,m_\pi^2,m_\pi^2;q^2) \, 
- 5/4\,m_\pi^4 \,H^{F\prime}(m_\pi^2,m_K^2,m_K^2;m_\pi^2) \,
\nonumber\\&&
+ 1/4\,m_\pi^2 \, H^F(m_\pi^2,m_K^2,m_K^2;m_\pi^2) \, 
- 1/4\,m_\pi^2 \, H^F(m_\pi^2,m_K^2,m_K^2;q^2) \, 
\nonumber\\&&
+ 1/9\,m_\pi^4 \, H^{F\prime}(m_\pi^2,m_\eta^2,m_\eta^2;m_\pi^2) \, 
+ 2\,m_\pi^2\,m_K^2 \, H^{F\prime}(m_K^2,m_\pi^2,m_K^2;m_\pi^2) \, 
\nonumber\\&&
- 2\,m_K^2 \, H^F(m_K^2,m_\pi^2,m_K^2;m_\pi^2) \, 
+ 2\,m_K^2 \, H^F(m_K^2,m_\pi^2,m_K^2;q^2) \, 
\nonumber\\&&
- 5/3\,m_\pi^4 \, H^{F\prime}(m_K^2,m_K^2,m_\eta^2;m_\pi^2) \, 
+  \bigg(
m_\pi^2 \, H^{F\prime}(m_\eta^2,m_K^2,m_K^2;m_\pi^2)
\nonumber\\&&
- H^F(m_\eta^2,m_K^2,m_K^2;m_\pi^2)
+ H^F(m_\eta^2,m_K^2,m_K^2;q^2) \bigg)\,(  - 1/4\,m_\pi^2 + m_K^2 )
\nonumber\\&&
+ 2\,m_\pi^4 \, H^{F\prime}_{1}(m_\pi^2,m_K^2,m_K^2;m_\pi^2) \, 
+ 4\,m_\pi^4 \, H^{F\prime}_{1}(m_K^2,m_K^2,m_\eta^2;m_\pi^2) \, 
\nonumber\\&&
+2\,H^T(m_\pi^2,m_\pi^2,m_\pi^2;q^2;m_\pi^2)
-1/4\, H^T((m_\pi^2,m_K^2,m_K^2;q^2;m_\pi^2)
\nonumber\\&&
+2 \,H^T(m_K^2,m_\pi^2,m_K^2;q^2;m_\pi^2)
+ 3/4\,H^T(m_\eta^2,m_K^2,m_K^2;q^2;m_\pi^2)\,.
\end{eqnarray}

The remainder for the kaon two-point function is
\begin{eqnarray}
\lefteqn{
\left(q^2 \hat\Pi_{AK}^{(1)}(q^2)\right)^{(6)}_{CT} =
  - 32\,m_\pi^2\,C^r_{13}\,q^2 - 32\,m_\pi^2\,C^r_{81}\,q^2 
 - 8\,m_K^4\,C^r_{91} - 32\,
  m_K^2\,C^r_{12}\,q^2
}&&\nonumber\\&&
 - 64\,m_K^2\,C^r_{13}\,q^2 - 32\,m_K^2\,C^r_{80}\,q^2 
 - 64\,m_K^2\,C^r_{81}\,q^2 - 16\,C^r_{87}\,q^4 - 8\,C^r_{93}\,q^4\,,
\end{eqnarray}
\begin{eqnarray}
\lefteqn{
\left(q^2 F_\pi^2\hat\Pi_{AK}^{(1)}(q^2)\right)^{(6)}_{loops} =
 \frac{1}{(16\pi^2)^2} \, \bigg( 1/24\,m_\pi^2\,q^2\,\pi^2 
 - 5/48\,m_\pi^2\,q^2 + 3/32\,m_K^4
}&&\nonumber\\&&
+ 5/24\,m_K^2\,q^2\,\pi^2 + 11/48\,m_K^2\,q^2 + 3/32\,q^4 \bigg)
+ \frac{1}{16\pi^2}\,\bigg( 3/4\,q^2 \mu_\pi \, 
\nonumber\\&&
+ 3/2\,q^2 \mu_K \, 
+ 3/4\,q^2 \mu_\eta \,\bigg)
+ 3/2\,m_\pi^{-2}\,q^2 \mu_\pi^2 
- 12\,L^r_{10}\,q^2\mu_\pi 
\nonumber\\&&
+ 3\,m_K^{-2}\,q^2 \mu_K^2  
- 24\,L^r_{10}\,q^2 \mu_K  
+ 3/2\,m_\eta^{-2}\,q^2 \mu_\eta^2 
- 12\,L^r_{10}\,q^2 \mu_\eta 
\nonumber\\&&
+ 3/4\,m_K^2 \,(2 \, m_\pi^2+ m_K^2 ) 
\, H^{F\prime}(m_\pi^2,m_\pi^2,m_K^2;m_K^2) \, 
\nonumber\\&&
 - 3/2\,m_\pi^2 \, H^F(m_\pi^2,m_\pi^2,m_K^2;m_K^2)
\nonumber\\&&
+ 3/2\,m_\pi^2 \, H^F(m_\pi^2,m_\pi^2,m_K^2;q^2) \, 
+ 1/2\,m_K^4 \, H^{F\prime}(m_\pi^2,m_K^2,m_\eta^2;m_K^2) \, 
\nonumber\\&&
- 3/16\,m_K^4 \, H^{F\prime}(m_K^2,m_\pi^2,m_\pi^2;m_K^2) \, 
+ 3/16\,m_K^2 \, H^F(m_K^2,m_\pi^2,m_\pi^2;m_K^2) \, 
\nonumber\\&&
- 3/16\,m_K^2 \, H^F(m_K^2,m_\pi^2,m_\pi^2;q^2) \, 
+ 9/8\,m_K^4 \, H^{F\prime}(m_K^2,m_\pi^2,m_\eta^2;m_K^2) \, 
\nonumber\\&&
- 9/8\,m_K^2\, H^F(m_K^2,m_\pi^2,m_\eta^2;m_K^2) \, 
+ 9/8\,m_K^2 \, H^F(m_K^2,m_\pi^2,m_\eta^2;q^2) \, 
\nonumber\\&&
+ 3/2\,m_K^4 \, H^{F\prime}(m_K^2,m_K^2,m_K^2;m_K^2) \, 
- 3/2\,m_K^2 \, H^F(m_K^2,m_K^2,m_K^2;m_K^2) \, 
\nonumber\\&&
+ 3/2\,m_K^2 \, H^F(m_K^2,m_K^2,m_K^2;q^2) \, 
+ 181/144\,m_K^4 \, H^{F\prime}(m_K^2,m_\eta^2,m_\eta^2;m_K^2) \, 
\nonumber\\&&
- 9/16\,m_K^2 \, H^F(m_K^2,m_\eta^2,m_\eta^2;m_K^2) \, 
+ 9/16\,m_K^2 \, H^F(m_K^2,m_\eta^2,m_\eta^2;q^2) \, 
\nonumber\\&&
- 3\,m_K^4 \, H^{F\prime}_{1}(m_\pi^2,m_\pi^2,m_K^2;m_K^2) \, 
- 3\,m_K^4 \, H^{F\prime}_{1}(m_K^2,m_\pi^2,m_\eta^2;m_K^2) \, 
\nonumber\\&&
- 5/2\,m_K^4 \, H^{F\prime}_{1}(m_K^2,m_\eta^2,m_\eta^2;m_K^2) \, 
+ 3/2 \,H^T(m_\pi^2,m_\pi^2,m_K^2;q^2;m_K^2)
\nonumber\\&&
-3/16\,H^T(m_K^2,m_\pi^2,m_\pi^2;q^2;m_K^2)
+ 9/8\,H^T(m_K^2,m_\pi^2,m_\eta^2;q^2;m_K^2)
\nonumber\\&&
+ 3/2\,H^T(m_K^2,m_K^2,m_K^2;q^2;m_K^2)
+ 9/16\,H^T(m_K^2,m_\eta^2,m_\eta^2;q^2;m_K^2)\,.
\end{eqnarray}

And finally the remainder for the hypercharge is
\begin{eqnarray}
\lefteqn{
\left(q^2 \hat\Pi_{A\eta}^{(1)}(q^2)\right)^{(6)}_{CT} =
256/3\,m_\pi^4\,C^r_{37} - 8\,m_\pi^4\,C^r_{91} 
 - 512/3\,m_\pi^2\,m_K^2\,C^r_{37}}
\nonumber\\&& 
+ 64/3\,m_\pi^2\,m_K^2\,C^r_{91}
+ 32/3\,m_\pi^2\,C^r_{12}\,q^2 - 32\,m_\pi^2\,C^r_{13}\,q^2 
 + 32/3\,m_\pi^2\,C^r_{80}\,q^2 
\nonumber\\&&
- 32\,m_\pi^2\,C^r_{81}\,q^2 
+ 256/3\,m_K^4\,C^r_{37} - 64/3\,m_K^4\,C^r_{91} 
 - 128/3\,m_K^2\,C^r_{12}\,q^2 
\nonumber\\&&
- 64\,m_K^2\,C^r_{13}\,q^2 
 - 128/3\,m_K^2\,C^r_{80}\,q^2 
 - 64\,m_K^2\,C^r_{81}\,q^2
   - 16\,C^r_{87}\,q^4 - 8\,C^r_{93}\,q^4 \,,
\nonumber\\
\end{eqnarray}
\begin{eqnarray}
\lefteqn{
\left(q^2 F_\pi^2\hat\Pi_{A\eta}^{(1)}(q^2)\right)^{(6)}_{loops} =
 \frac{1}{(16\pi^2)^2} \, 
\bigg( 1/96\,m_\pi^4 - 1/12\,m_\pi^2\,m_K^2- 1/4\,m_\pi^2\,q^2
}&&\nonumber\\&&
 + 1/6\,m_K^4
+ 1/4\,m_K^2\,q^2\,\pi^2 + 3/8\,m_K^2\,q^2 + 3/32\,q^4 \bigg)
+ \frac{1}{16\pi^2}\,3\,q^2 \, \mu_K  
\nonumber\\&&
+ 6\,m_K^{-2}\,q^2 \, \mu_K^2  
- 48\,L^r_{10}\,q^2 \,\mu_K \, 
+ 1/3\,m_\pi^4 \, H^{F\prime}(m_\pi^2,m_\pi^2,m_\eta^2;m_\eta^2) \, 
\nonumber\\&&
+ ( m_\pi^4 + 9/4\,m_\pi^2\,m_\eta^2 ) \, 
H^{F\prime}(m_\pi^2,m_K^2,m_K^2;m_\eta^2) 
\nonumber\\&&
- 9/4\,m_\pi^2 \, H^F(m_\pi^2,m_K^2,m_K^2;m_\eta^2) \, 
+ 9/4\,m_\pi^2 \, H^F(m_\pi^2,m_K^2,m_K^2;q^2) \, 
\nonumber\\&&
+ (  -m_\pi^4 + 8\,m_\pi^2\,m_K^2 - 128/9\, m_K^4 + 9/4\,m_\eta^4 )
\, H^{F\prime}(m_K^2,m_K^2,m_\eta^2;m_\eta^2) \, 
\nonumber\\&&
 -9/4\,m_\eta^2 \,\bigg( H^F(m_\eta^2,m_K^2,m_K^2;m_\eta^2)
- H^F(m_\eta^2,m_K^2,m_K^2;q^2) \bigg) 
\nonumber\\&&
+ ( 49/243\,m_\pi^4 - 224/243\,m_\pi^2\, m_K^2 + 256/243\,m_K^4 )\,
H^{F\prime}(m_\eta^2,m_\eta^2,m_\eta^2;m_\eta^2) \, 
\nonumber\\&&
-6\,m_\pi^2 m_\eta^2 \, H^{F\prime}_{1}(m_\pi^2,m_K^2,m_K^2;m_\eta^2) \, 
\nonumber\\&&
+ 4\,(-3\,m_\pi^2 + 8\,\,m_K^2 )\,m_\eta^2 
\, H^{F\prime}_{1}(m_K^2,m_K^2,m_\eta^2;m_\eta^2) \, 
\nonumber\\&&
+9/4\, H^T(m_\pi^2,m_K^2,m_K^2;q^2;m_\eta^2)
+9/4\,H^T(m_\eta^2,m_K^2,m_K^2;q^2;m_\eta^2)\,.
\end{eqnarray}

\subsection{The remainder longitudinal part}
\label{longitudinal}

In that case the $\pfour$~ contribution vanishes so we have that
\begin{equation}
\left.\hat\Pi_{Aa}^{(0)}(q^2) =
\left(\hat\Pi_{Aa}^{(0)}(q^2)\right)^{(6)}_{CT}
+\left(\hat\Pi_{Aa}^{(0)}(q^2)\right)^{(6)}_{loops}
\right|_{a = \pi,K,\eta}\;.
\end{equation}

The expansion of the
resummed selfenergy around the relevant pseudoscalar mass leads
in general to rather high derivatives and produces
naturally the combinations
\begin{eqnarray}
\lefteqn{
H^L_i(m_1^2,m_2^2,m_3^2;q^2;m_4^2) =
\frac{1}{(q^2-m_4^2)^2}\bigg(
q^2 H^F_i(m_1^2,m_2^2,m_3^2;q^2)
}&&\nonumber\\&&
- q^2 H^F_i(m_1^2,m_2^2,m_3^2;m_4^2)
-m_4^2(q^2-m_4^2)H^{F\prime}_i(m_1^2,m_2^2,m_3^2;m_4^2)\bigg)
\end{eqnarray}
for $H_i^F = \{H^F,H^F_1,H^F_{21}\}$ and
\begin{eqnarray}
\lefteqn{
H^M(m_1^2,m_2^2,m_3^2;q^2;m_4^2) =
\frac{1}{(q^2-m_4^2)^2}\bigg(
H^F(m_1^2,m_2^2,m_3^2;q^2)
}&&\nonumber\\&&
-H^F(m_1^2,m_2^2,m_3^2;m_4^2)
-(q^2-m_4^2)H^{F\prime}(m_1^2,m_2^2,m_3^2;m_4^2)\bigg)\,.
\end{eqnarray}
All of these functions are
regular at $q^2 = m_4^2$.
 
The longitudinal isospin remainder is
\begin{equation}
\left(q^2\hat\Pi_{A\pi}^{(0)}(q^2)\right)^{(6)}_{CT} =
   8\,m_\pi^4\,C^r_{91} \,,
\end{equation}
\begin{eqnarray}
\lefteqn{
\left(q^2 F_\pi^2\hat\Pi_{A\pi}^{(0)}(q^2)\right)^{(6)}_{loops} =
\frac{- 3/32\,m_\pi^4}{(16\pi^2)^2} \, 
}&&\nonumber\\&&
+ 1/3\,H^L(m_\pi^2,m_\pi^2,m_\pi^2;q^2;m_\pi^2)\,m_\pi^4 
- 2\,H^M(m_\pi^2,m_\pi^2,m_\pi^2;q^2;m_\pi^2)\,m_\pi^6 
\nonumber\\&& 
+      H^L(m_\pi^2,m_K^2,m_K^2;q^2;m_\pi^2)\,m_\pi^4
+ 1/4\,H^M(m_\pi^2,m_K^2,m_K^2;q^2;m_\pi^2)\,m_\pi^6 
\nonumber\\&& 
- 1/9\,H^L(m_\pi^2,m_\eta^2,m_\eta^2;q^2;m_\pi^2)\,m_\pi^4 
- 2\,H^M(m_K^2,m_\pi^2,m_K^2;q^2;m_\pi^2)\,m_\pi^4\,m_K^2 
\nonumber\\&& 
+ 5/3\,H^L(m_K^2,m_K^2,m_\eta^2;q^2;m_\pi^2)\,m_\pi^4 
- 3/4\,H^M(m_\eta^2,m_K^2,m_K^2;q^2;m_\pi^2)\,m_\pi^4 m_\eta^2
\nonumber\\&& 
- 2\,H^L_{1}(m_\pi^2,m_K^2,m_K^2;q^2;m_\pi^2)\,m_\pi^4 
- 4\,H^L_{1}(m_K^2,m_K^2,m_\eta^2;q^2;m_\pi^2)\,m_\pi^4 
\nonumber\\&& 
- 6\,H^L_{21}(m_\pi^2,m_\pi^2,m_\pi^2;q^2;m_\pi^2)\,m_\pi^4 
+ 3/4\,H^L_{21}(m_\pi^2,m_K^2,m_K^2;q^2;m_\pi^2)\,m_\pi^4 
\nonumber\\&& 
- 6\,H^L_{21}(m_K^2,m_\pi^2,m_K^2;q^2;m_\pi^2)\,m_\pi^4 
- 9/4\,H^L_{21}(m_\eta^2,m_K^2,m_K^2;q^2;m_\pi^2)\,m_\pi^4 \,.
\end{eqnarray}

The kaon  remainder is
\begin{equation}
\left(q^2 \hat\Pi_{AK}^{(0)}(q^2)\right)^{(6)}_{CT} =
 8\,m_K^4\,C^r_{91} \,,
\end{equation}
\begin{eqnarray}
\lefteqn{
\left(q^2 F_\pi^2\hat\Pi_{AK}^{(0)}(q^2)\right)^{(6)}_{loops} =
 \frac{- 3/32\,m_K^4}{(16\pi^2)^2} \,  
}&&\nonumber\\&&
- 3/2\,H^M(m_\pi^2,m_\pi^2,m_K^2;q^2;m_K^2)\,m_\pi^2\,m_K^4 
+ 3/16\,H^M(m_K^2,m_\pi^2,m_\pi^2;q^2;m_K^2)\,m_K^6 
\nonumber\\&&
- 3/4\,H^L(m_\pi^2,m_\pi^2,m_K^2;q^2;m_K^2)\,m_K^4 
- 1/2\,H^L(m_\pi^2,m_K^2,m_\eta^2;q^2;m_K^2)\,m_K^4 
\nonumber\\&&
- 9/8\,H^M(m_K^2,m_\pi^2,m_\eta^2;q^2;m_K^2)\,m_K^6 
- 3/2\,H^M(m_K^2,m_K^2,m_K^2;q^2;m_K^2)\,m_K^6 
\nonumber\\&&
  - 25/36\,H^L(m_K^2,m_\eta^2,m_\eta^2;q^2;m_K^2)\,m_K^4
- 9/16\,H^M(m_K^2,m_\eta^2,m_\eta^2;q^2;m_K^2)\,m_K^6 
\nonumber\\&&
+ 3\,H^L_{1}(m_\pi^2,m_\pi^2,m_K^2;q^2;m_K^2)\,m_K^4 
+ 3\,H^L_{1}(m_K^2,m_\pi^2,m_\eta^2;q^2;m_K^2)\,m_K^4 
\nonumber\\&&
+ 5/2\,H^L_{1}(m_K^2,m_\eta^2,m_\eta^2;q^2;m_K^2)\,m_K^4 
- 9/2\,H^L_{21}(m_\pi^2,m_\pi^2,m_K^2;q^2;m_K^2)\,m_K^4 
\nonumber\\&&
+ 9/16\,H^L_{21}(m_K^2,m_\pi^2,m_\pi^2;q^2;m_K^2)\,m_K^4 
- 27/8\,H^L_{21}(m_K^2,m_\pi^2,m_\eta^2;q^2;m_K^2)\,m_K^4 
\nonumber\\&&
- 9/2\,H^L_{21}(m_K^2,m_K^2,m_K^2;q^2;m_K^2)\,m_K^4 
- 27/16\,H^L_{21}(m_K^2,m_\eta^2,m_\eta^2;q^2;m_K^2)\,m_K^4\,.
\nonumber\\ 
\end{eqnarray}

Finally the hypercharge remainder is
\begin{eqnarray}
\lefteqn{
\left(q^2 \hat\Pi_{A\eta}^{(0)}(q^2)\right)^{(6)}_{CT} =
  - 256/3\,m_\pi^4\,C^r_{37} + 8\,m_\pi^4\,C^r_{91} 
    + 512/3\,m_\pi^2\,m_K^2\,C^r_{37}
}&&\nonumber\\&&   
- 64/3\,m_\pi^2\,m_K^2\,C^r_{91} - 256/3\,m_K^4\,C^r_{37} 
 + 64/3\,m_K^4\,C^r_{91}\,, \hspace{2cm}
\end{eqnarray}
\begin{eqnarray}
\lefteqn{
\left(q^2 F_\pi^2\hat\Pi_{A\eta}^{(0)}(q^2)\right)^{(6)}_{loops} =
 \frac{1}{(16\pi^2)^2} \, 
\bigg(  - 1/96\,m_\pi^4 + 1/12\,m_\pi^2\,m_K^2 - 1/6\,m_K^4 \bigg)
}&&\nonumber\\&&
- 1/3\,H^L(m_\pi^2,m_\pi^2,m_\eta^2;q^2;m_\eta^2)\,m_\pi^4 
- H^L(m_\pi^2,m_K^2,m_K^2;q^2;m_\eta^2)\,m_\pi^4 
\nonumber\\&&
- 9/4\,H^M(m_\pi^2,m_K^2,m_K^2;q^2;m_\eta^2)\,m_\pi^2 m_\eta^4 
\nonumber\\&&
+ H^L(m_K^2,m_K^2,m_\eta^2;q^2;m_\eta^2)\,
  (m_\pi^4 - 8\,m_\pi^2\,m_K^2+ 128/9\,m_K^4 )
\nonumber\\&&
- 9/4 H^M(m_\eta^2,m_K^2,m_K^2;q^2;m_\eta^2)\,m_\eta^6
\nonumber\\&&
+ 1/243\,H^L(m_\eta^2,m_\eta^2,m_\eta^2;q^2;m_\eta^2)
   (- 49\,m_\pi^4 + 224\,m_\pi^2\,m_K^2 - 256\,m_K^4 )
\nonumber\\&&
+ 6H^L_1(m_\pi^2,m_K^2,m_K^2;q^2;m_\eta^2) m_\pi^2 m_\eta^2
+ H^L_1(m_K^2,m_K^2,m_\eta^2;q^2;m_\eta^2)4(3 m_\pi^2-8 m_K^2)m_\eta^2
\nonumber\\&&
-27/4\,H^L_{21}(m_\pi^2,m_K^2,m_K^2;q^2;m_\eta^2) m_\eta^4
-27/4\,H^L_{21}(m_\eta^2,m_K^2,m_K^2;q^2;m_\eta^2) m_\eta^4\,.
\end{eqnarray}

\section{Loop integrals}
\label{loopint}

We use dimensional regularization here throughout in $d$ dimensions
with $d=4-2\epsilon$.

\subsection{One-loop integrals}
\label{oneloopint}

We need integrals with one, two and three propagators in principle.
These we define by
\be
A(m^2) = \frac{1}{i}\int \frac{d^d q}{(2\pi)^d}\frac{1}{q^2-m^2}\,.
\ee
We also use below
\be
A(n,m^2) = \frac{1}{i}\int \frac{d^d q}{(2\pi)^d}
\frac{1}{\left(q^2-m^2\right)^n}\,,
\ee
which can be obtained by derivation w.r.t. $m^2$ of $A(m^2)$.

The two propagator integrals we encounter are
\ba
B(m_1^2,m_2^2,p^2)&=& \frac{1}{i}\int\frac{d^dq}{(2\pi)^d}
\frac{1}{(q^2-m_1^2)((q-p)^2-m_2^2)}\,,
\nonumber\\
B_\mu(m_1^2,m_2^2,p)
&=& \frac{1}{i}\int\frac{d^dq}{(2\pi)^d}
\frac{q_\mu}{(q^2-m_1^2)((q-p)^2-m_2^2)}
\nonumber\\
&=& p_\mu B_1(m_1^2,m_2^2,p^2)\,,
\nonumber\\
B_{\mu\nu}(m_1^2,m_2^2,p)
&=& \frac{1}{i}\int\frac{d^dq}{(2\pi)^d}
\frac{q_\mu q_\nu}{(q^2-m_1^2)((q-p)^2-m_2^2)}
\nonumber\\
&=& p_\mu p_\nu B_{21}(m_1^2,m_2^2,p^2)
+ g_{\mu\nu} B_{22}(m_1^2,m_2^2,p^2)\,.
\ea

All the cases with three propagator integrals that show up can be absorbed
into the two-propagator ones by moving to the real masses rather than the
lowest order masses. This provided in fact a consistency check on the
calculations.

The explicit expressions are well known
\ba
A(m^2) & = &\frac{m^2}{16\pi^2}\Bigg\{\lambda_0-\ln(m^2)
+\epsilon\Big(\frac{C^2}{2}+\frac{1}{2}+\frac{\pi^2}{12}
+\frac{1}{2}\ln^2(m^2) 
\nonumber \\& &
- C\ln(m^2)\Big)\Bigg\}
+{\cal O}(\epsilon^2)
\,,
\nonumber\\
B(m_1^2,m_2^2,p^2) &=&
\frac{1}{16\pi^2}\left(\lambda_0
-\frac{m_1^2\ln(m_1^2)-m_2^2\ln(m_2^2)}{m_1^2-m_2^2}\right)
+\bar{J}(m_1^2,m_2^2,p^2)+{\cal O}(\epsilon)\,,
\nonumber\\
\bar{J}(m_1^2,m_2^2,p^2) &=&
-\frac{1}{16\pi^2}\int_0^1 dx
\ln\left(\frac{m_1^2 x+m_2^2(1-x)-x(1-x)p^2}{m_1^2 x+m_2^2(1-x)}\right)\,,
\ea
$C=\ln(4\pi)+1-\gamma$ and $\lambda_0 = 1/\epsilon+C$.
The function $\bar{J}(m_1^2,m_2^2,p^2)$ develops an imaginary
part for $p^2\ge (m_1+m_2)^2$. Using $\Delta=m_1^2-m_2^2$, $\Sigma=m_1^2+m_2^2$
and $\nu^2 = p^4+m_1^4+m_2^4-2p^2m_1^2-2p^2m_2^2-2m_1^2m_2^2$
it is given by
\ba
(32\pi^2)\bar{J}(m_1^2,m_2^2,p^2) &=&
2+\left(-\frac{\Delta}{p^2}+\frac{\Sigma}{\Delta}\right)\ln\frac{m_1^2}{m_2^2}
-\frac{\nu}{p^2}\ln\frac{(p^2+\nu)^2-\Delta^2}{(p^2-\nu)^2-\Delta^2}\,.
\ea

The two-propagator integrals can all be reduced to $B$ and $A$ via
\ba
B_1(m_1^2,m_2^2,p^2)& = &-\frac{1}{2p^2}\left(
  A(m_1^2)-A(m_2^2)+(m_2^2-m_1^2-p^2) B(m_1^2,m_2^2,p^2)\right)\,,
\nonumber\\
B_{22}(m_1^2,m_2^2,p^2)& = & \frac{1}{2(d-1)}\Big(
 A(m_2^2)+2 m_1^2 B(m_1^2,m_2^2,p^2)
\nonumber\\&&
+(m_2^2-m_1^2-p^2) B_1(m_1^2,m_2^2,p^2)
\Big)\,,
\nonumber\\
B_{21}(m_1^2,m_2^2,p^2)& = &\frac{1}{p^2}\left(
  A(m_2^2)+m_1^2 B(m_1^2,m_2^2,p^2)-d B_{22}(m_1^2,m_2^2,p^2)\right)\,.
\ea
The basic method used here is the one from Passarino and Veltman \cite{PV}.

\subsection{Sunset Integrals}

In this appendix we discuss the nontrivial two-loop
integrals that show up in this calculation.
They have been treated in several places already, in general and for
various special cases. We use here a method that is a hybrid of
various other approaches. We only cite the literature actually used.
We define
\newcommand{\lla}{\langle\langle}
\newcommand{\rra}{\rangle\rangle}
\be
\lla X \rra 
= \frac{1}{i^2}\int \frac{d^d q}{(2\pi)^d} \frac{d^d r}{(2\pi)^d}
\frac{X}
{\left(q^2-m_1^2\right)\left(r^2-m_2^2\right)\left((q+r-p)^2-m_3^2\right)}\,,
\ee
\ba
\label{defsunset}
H(m_1^2,m_2^2,m_3^2;p^2) &=&\lla 1\rra\,,\nonumber\\
H_\mu(m_1^2,m_2^2,m_3^2;p^2) &=&\lla q_\mu\rra
= p_\mu H_1(m_1^2,m_2^2,m_3^2;p^2)\,,\nonumber\\
H_{\mu\nu}(m_1^2,m_2^2,m_3^2;p^2) &=&\lla q_\mu q_\nu\rra\nonumber\\
& = &
p_\mu p_\nu H_{21}(m_1^2,m_2^2,m_3^2;p^2)
+g_{\mu\nu} H_{22}(m_1^2,m_2^2,m_3^2;p^2)
\,.
\nonumber\\
\ea
By redefining momenta the others can
be simply related to the above three.
In particular
\ba
\lla r_\mu\rra &=& p_\mu H_1(m_2^2,m_1^2,m_3^2;p^2)\,,\nonumber\\
\lla r_\mu r_\nu\rra &=& p_\mu p_\nu H_{21}(m_2^2,m_1^2,m_3^2;p^2)
+ g_{\mu\nu} H_{22}(m_2^2,m_1^2,m_3^2;p^2)\,,\nonumber\\
\lla q_\mu r_\nu \rra &=& \lla r_\mu q_\nu\rra \,,\nonumber\\
\lla q_\mu r_\nu\rra &=&p_\mu p_\nu H_{23}(m_1^2,m_2^2,m_3^2;p^2)
+g_{\mu\nu} H_{24}(m_1^2,m_2^2,m_3^2;p^2)\,,
\ea
with
\ba
2H_{23}(m_1^2,m_2^2,m_3^2;p^2)&=&
-H_{21}(m_1^2,m_2^2,m_3^2;p^2)
-H_{21}(m_2^2,m_1^2,m_3^2;p^2)
\nonumber\\&&
+H_{21}(m_3^2,m_1^2,m_2^2;p^2)
+2 H_1(m_1^2,m_2^2,m_3^2;p^2)
\nonumber\\&&
+2 H_1(m_2^2,m_1^2,m_3^2;p^2)
-H(m_1^2,m_2^2,m_3^2;p^2)\,,
\nonumber\\
2 H_{24}(m_1^2,m_2^2,m_3^2;p^2)&=&
-H_{22}(m_1^2,m_2^2,m_3^2;p^2)
-H_{22}(m_2^2,m_1^2,m_3^2;p^2)
\nonumber\\&&
+H_{22}(m_3^2,m_1^2,m_2^2;p^2)\,.
\ea
The first two follow from interchanging $q$ and $r$ and the third
from the fact that it is proportional to $g_{\mu\nu}$ or $p_\mu p_\nu$, which
are both symmetric in $\mu$ and $\nu$. The last one follows from
\ba
\left(q_\mu r_\nu+r_\mu q_\nu\right) &=&
\left(q_\mu+r_\mu-p_\mu\right)\left(q_\nu+r_\nu-p_\nu\right)
-q_\mu q_\nu - r_\mu r_\nu -p_\mu p_\nu 
\nonumber\\&&+2p_\mu\left(q_\nu+r_\nu\right)
+2 p_\nu\left(q_\mu+r_\mu\right)
\ea
and redefining momenta and masses on the r.h.s..
In addition we have the relation
\ba
\label{H22relation}
\lefteqn{
p^2 H_{21}(m_1^2,m_2^2,m_3^2;p^2)+d H_{22}(m_1^2,m_2^2,m_3^2;p^2)
= }&&
\nonumber\\&&
 m_1^2 H(m_1^2,m_2^2,m_3^2;p^2)+A(m_2^2) A(m_3^2)\,.
\ea
which allows to express $H_{22}$ in a simple way in terms
of $H_{21}$.
There is also a relation between $H_1$ and $H$
\ba
 H_1(m_1^2,m_2^2,m_3^2;p^2)+ H_1(m_2^2,m_1^2,m_3^2;p^2)
 + H_1(m_3^2,m_1^2,m_2^2;p^2)= 
\nonumber\\
H(m_1^2,m_2^2,m_3^2;p^2)\,,
\ea
which  allows to write $H_1(m^2,m^2,m^2;p^2)
= 1/3\;H(m^2,m^2,m^2;p^2)$ in the case of equal masses.
The function $H$ is fully symmetric in $m_1^2,m_2^2$ and $m_3^2$, while
$H_1$, $H_{21}$ and $H_{22}$ are symmetric under the interchange
of $m_2^2$ and $m_3^2$.

We do not explicitly evaluate the integrals analytically.
$H$, $H_1$ and $H_{21}$ are all finite after two subtractions.
We therefore evaluate them as follows ($H_i$ stands for
$H$, $H_1$ and $H_{21}$)
\ba
\label{expandHi}
H_i(m_1^2,m_2^2,m_3^2;p^2) &=& H_i(m_1^2,m_2^2,m_3^2;0)
+ p^2 \frac{\partial}{\partial p^2}H_i(m_1^2,m_2^2,m_3^2;0)
\nonumber\\&&
+\overline{H}_i(m_1^2,m_2^2,m_3^2;p^2)\;.
\ea
The functions $\overline{H}_i$
are finite in 4 dimensions and can be evaluated by their dispersive
representation \cite{PT,GS} or below threshold by the methods of \cite{GS}.

The value at zero and its derivative there have been derived essentially using the
methods of \cite{DT} except that we use a slightly simpler procedure
than the recursion relations given there.

First we define the intermediate integrals
\be
I(n_1,n_2,n_3) =  
\frac{1}{i^2}\int \frac{d^d q}{(2\pi)^d} \frac{d^d r}{(2\pi)^d}
\frac{1}
{\left(q^2-m_1^2\right)^{n_1}
\left(r^2-m_2^2\right)^{n_2}\left((q+r)^2-m_3^2\right)^{n_3}}\,,
\ee
which show up in the momentum expansion of $H_i$.
The $I(n_1,n_2,n_3)$ with one of the $n_i=0$ are separable and are
e.g.
\be
I(n_1,n_2,0) = A(n_1,m_1^2) A(n_2,m_2^2)\,.
\ee
All the others can be derived by taking derivatives of $I(1,1,1)$ w.r.t.
the masses $m_1^2,m_2^2$ and $m_3^2$. The function $I(1,1,1)$
is taken from \cite{DT}, note that our definition differs by overall factors
from theirs.
\ba
\lefteqn{\label{I111}I(1,1,1) = \frac{-1}{(16\pi^2)^2}
\frac{\Gamma^2(1+\epsilon)}{2(1-\epsilon)(1-2\epsilon)}
(4\pi)^{2\epsilon} \Bigg\{ }
&&\nonumber\\
&&
-\frac{1}{\epsilon^2}\left(m_1^2+m_2^2+m_3^2\right) 
+\frac{2}{\epsilon}\left(m_1^2\ln_1+m_2^2\ln_2+m_3^2\ln_3\right)
\nonumber\\&&
+m_1^2\;\left(\ln_2\ln_3-\ln_1(\ln_1+\ln_2+\ln_3)\right)
+m_2^2\;\left(\ln_3\ln_1-\ln_2(\ln_1+\ln_2+\ln_3)\right)
\nonumber\\&&
+m_3^2\;\left(\ln_1\ln_2-\ln_3(\ln_1+\ln_2+\ln_3)\right)
+\Psi(m_1^2,m_2^2,m_3^2)
+{\cal O}(\epsilon)
\Bigg\}\,.
\ea
In (\ref{I111}) we used $\ln_i = \ln(m_i^2)$ 
and the function
$\Psi(m_1^2,m_2^2,m_3^2)$. The expression for $\Psi$ is somewhat dependent
on the relation between the various masses. Using
\be
\lambda_m =\lambda(m_1^2,m_2^2,m_3^2) = (m_1^2-m_2^2-m_3^2)^2-4 m_2^2 m_3^2\,,
\ee
we have for the case $\lambda_m\le 0$\cite{DT}
\newcommand{\cl}{\mathrm{Cl}_2}
\newcommand{\dilog}{\mathrm{Li}_2}
\be
\Psi(m_1^2,m_2^2,m_3^2) = 2\sqrt{-\lambda_m}
\left\{\cl(2\arccos z_1)+\cl(2\arccos z_2)+\cl(2\arccos z_3)\right\}\,,
\ee
with
\be
z_1 = \frac{-m_1^2+m_2^2+m_3^2}{2 m_2 m_3}\,,\quad
z_2 = \frac{-m_2^2+m_3^2+m_1^2}{2 m_3 m_1}\,,\quad
z_3 = \frac{-m_3^2+m_1^2+m_2^2}{2 m_1 m_2}\,.
\ee
The case $m_1+m_2\le m_3$, with $\lambda_m\ge0$, is
\ba
\Psi(m_1^2,m_2^2,m_3^2) &=& -\sqrt{\lambda_m}
\Bigg\{2\ln x_1 \ln x_2-\ln\frac{m_1^2}{m_3^2}\ln\frac{m_2^2}{m_3^2}
\nonumber\\&&
+\frac{\pi^2}{3}-2\dilog(x_1)-2\dilog(x_2)
\Bigg\}\,,
\ea
with
\be
x_1 = \frac{m_3^2+m_1^2-m_2^2-\sqrt{\lambda_m}}{2m_3^2}\,,\quad
x_2 = \frac{m_3^2+m_2^2-m_1^2-\sqrt{\lambda_m}}{2m_3^2}\,.
\ee
The cases $m_1+m_3\le m_2$ and $m_2+m_3\le m_1$ can be obtained from the last
one by relabelling masses.
$\dilog(x)$ is the dilogarithm defined by
\be
\dilog(x) = -\int_0^1 \frac{dt}{t}\ln(1-xt)\,,
\ee
and $\cl(x)$ is Clausen's function defined by
\be
\cl(x) = -\int_0^x dt \ln \vert 2\sin\frac{t}{2}\vert
= -i\left(\dilog(e^{ix})-\dilog(1)\right)
-i\frac{\pi x}{2}+i\frac{x^2}{4}\,.
\ee

Notice that $\Psi(m_1^2,m_2^2,m_3^2)$ is fully symmetric w.r.t. the three
masses.
The $I(n_1,n_2,n_3)$ for general $n_i$ can be obtained by taking derivatives
of $I(1,1,1)$. The relation
\be
\frac{\partial}{\partial m_1^2}\Psi(m_1^2,m_2^2,m_3^2) =
\frac{m_1^2-m_2^2-m_3^2}{\lambda_m}\Psi(m_1^2,m_2^2,m_3^2)
-\ln\frac{m_1^4}{m_2^2 m_3^2}
+\frac{m_2^2-m_3^2}{m_1^2}\ln\frac{m_2^2}{m_3^2}\,,
\ee
allows an easy evaluation of all needed derivatives and is equivalent to
the recursion relations used in \cite{DT}.

In order to express the $H_i$ functions at zero and the derivatives
w.r.t. $p^2$ at zero the easiest is to shift momenta to
$\tilde q=q-p$ in the integral and then Taylor-expand using
\be
\frac{1}{(\tilde q+p)^2-m_1^2} = \frac{1}{\tilde q^2-m_1^2}\,
\sum_{i=0,\infty}
\left(\frac{-2\tilde q\cdot p-p^2}{\tilde q^2-m_1^2}\right)^i \;.
\ee
The integrals can then be done using
$\tilde q_\mu \tilde q_\nu\longrightarrow g_{\mu\nu}q^2/d$ and equivalent
identities for the higher orders. We have run this procedure to higher orders
then necessary to check the cancellations of infinities there.
This results in
\ba
H(m_1^2,m_2^2,m_3^2;0) &=& I(1,1,1)\,,
\nonumber\\
\frac{\partial}{\partial p^2}H(m_1^2,m_2^2,m_3^2;0) &=& 
\frac{4-d}{d}I(2,1,1)+\frac{4}{d}m_1^2 I(3,1,1)\,,
\nonumber\\
H_1(m_1^2,m_2^2,m_3^2;0) &=& \frac{d-2}{d}I(1,1,1)-\frac{2}{d}m_1^2 I(2,1,1))\,,
\nonumber\\ 
\frac{\partial}{\partial p^2}H_1(m_1^2,m_2^2,m_3^2;0) &=&
\frac{1}{d(d+2)} \Big((2-d)(d-4) I(2,1,1)
\nonumber\\&&
+8(d-4)m_1^2 I(3,1,1)
-24 m_1^4 I(4,1,1)\Big)\,,
\nonumber\\
H_{21}(m_1^2,m_2^2,m_3^2;0) &=& \frac{1}{d(d+2)}
\Big(d(d-2)I(1,1,1)-4(d-2) m_1^2 I(2,1,1)
\nonumber\\&&
+8 m_1^4 I(3,1,1)\Big)\,,
\nonumber\\
\frac{\partial}{\partial p^2}H_{21}(m_1^2,m_2^2,m_3^2;0) &=&
\frac{1}{d(d+2)}\Big\{(-d^2+10d-48)I(2,1,1)
\nonumber\\&&
\hskip-4cm +12(d-10)m_1^2 I(3,1,1)-72m_1^4 I(4,1,1)
+\frac{192}{d+4}\Big(
I(2,1,1)
\nonumber\\&&
\hskip-4cm +3m_1^2 I(3,1,1)
+3 m_1^4 I(4,1,1)+m_1^6 I(5,1,1)\Big)\Big\}\,.
\ea
Evaluating these expressions then leads to
\ba
\label{H}
\lefteqn{
(16\pi^2)^2 H(m_1^2,m_2^2,m_3^2;0) = 
\frac{1}{2}\lambda_2 (m_1^2+m_2^2+m_3^2)
+\frac{1}{2}\lambda_1\Big(m_1^2(1-2\ln_1)
}&&\nonumber\\&&
+m_2^2 (1-2\ln_2)+m_3^2 (1-2\ln_3)\Big)
-\frac{1}{2}\Psi(m_1^2,m_2^2,m_3^2)
+m_1^2\Bigg(\frac{\pi^2}{12}+\frac{3}{2}-\ln_1
\nonumber\\&&
+\frac{1}{2}
\Big(-\ln_2\ln_3+\ln_1\ln_4\Big)\Bigg)
+m_2^2\Bigg(\frac{\pi^2}{12}+\frac{3}{2}-\ln_2+\frac{1}{2}
\Big(-\ln_1\ln_3+\ln_2\ln_4\Big)\Bigg)
\nonumber\\&&
+m_3^2\Bigg(\frac{\pi^2}{12}+\frac{3}{2}-\ln_3+\frac{1}{2}
\Big(-\ln_1\ln_2+\ln_3\ln_4\Big)\Bigg)+\cal{O}(\epsilon)\,,
\ea
\ba
\lefteqn{
(16\pi^2)^2
\frac{\partial}{\partial p^2}H(m_1^2,m_2^2,m_3^2;0) =
-\frac{1}{4}\lambda_1
+\frac{m_1^2 m_2^2 m_3^3}{\lambda_m^2}\Psi(m_1^2,m_2^2,m_3^2)+\frac{1}{8}
}&&\nonumber\\&&
+\frac{m_1^2 \delta_m}{2\lambda_m}\ln_1
+\frac{m_2^2}{2\lambda_m}(m_2^2-m_1^2-m_3^2)\ln_2
+\frac{m_3^2}{2\lambda_m}(m_3^2-m_1^2-m_2^2)\ln_3+\cal{O}(\epsilon)\,,
\nonumber\\&&
\ea
\ba
\lefteqn{
(16\pi^2)^2 H_1(m_1^2,m_2^2,m_3^2;0) = 
\frac{m_2^2+m_3^2}{4}\lambda_2+\frac{1}{8}\lambda_1\Big(2 m_1^2+m_2^2(1-4\ln_2)
}&&
\nonumber\\&&
+m_3^2 (1-4\ln_3)\Big)
+\frac{1}{4}\Bigg(-1+\frac{m_1^2\delta_m}{\lambda_m}\Bigg)
\Psi(m_1^2,m_2^2,m_3^2)
+m_1^2\Bigg(\frac{3}{8}-\frac{1}{2}\ln_1\Bigg)
\nonumber\\&&
+m_2^2\Bigg(\frac{\pi^2}{24}+\frac{9}{16}-\frac{1}{4}\ln_2
  +\frac{1}{4}(\ln_2\ln_4-\ln_1\ln_3)\Bigg)
\nonumber\\&&
+m_3^2\Bigg(\frac{\pi^2}{24}+\frac{9}{16}-\frac{1}{4}\ln_3
  +\frac{1}{4}(\ln_3\ln_4-\ln_1\ln_2)\Bigg)+\cal{O}(\epsilon)\,,
\ea
\ba
\lefteqn{
(16\pi^2)^2\frac{\partial}{\partial p^2}H_1(m_1^2,m_2^2,m_3^2;0) =
-\frac{1}{12}\lambda_1
+\frac{m_1^4m_2^2m_3^2\delta_m}{\lambda_m^3}\Psi(m_1^2,m_2^2,m_3^2)
-\frac{m_1^2\delta_m}{6\lambda_m}
}&&\nonumber\\&&
+\frac{m_1^4\ln_1}{6\lambda_m}\Bigg(1+12\frac{m_2^2m_3^2}{\lambda_m}\Bigg)
+\frac{m_2^2\ln_2}{6\lambda_m}\Bigg(m_2^2-m_3^2-2m_1^2
 - \frac{6m_1^2m_3^2}{\lambda_m}(\delta_m+2m_2^2)\Bigg)
\nonumber\\&&
+\frac{m_3^2\ln_3}{6\lambda_m}\Bigg(m_3^2-m_2^2-2m_1^2
 - \frac{6m_1^2m_2^2}{\lambda_m}(\delta_m+2m_3^2)\Bigg)
+\frac{7}{72}+\cal{O}(\epsilon)\,,
\ea
\ba
\lefteqn{
(16\pi^2)^2 H_{21}(m_1^2,m_2^2,m_3^2;0) = 
\frac{m_2^2+m_3^2}{6}\lambda_2+\frac{1}{36}\lambda_1\Bigg(
 3m_1^2+m_2^2(2-12\ln_2)
}&&\nonumber\\&&
+m_3^2(2-12\ln_3)\Bigg)
+\frac{1}{6\lambda_m^2}\Bigg(-\lambda_m^2+\lambda_m m_1^2\delta_m
   +2m_1^4m_2^2m_3^2\Bigg)\Psi(m_1^2,m_2^2,m_3^2)
\nonumber\\&&
+m_1^2 \Bigg(\frac{17}{72}-\frac{1}{3}\ln_1
  +\frac{m_1^2\ln_1\delta_m}{6\lambda_m}\Bigg)
\nonumber\\&&
+m_2^2\Bigg(\frac{\pi^2}{36}+\frac{19}{54}-\frac{1}{9}\ln_2
  -\frac{m_1^2}{6\lambda_m}(\delta_m+2m_3^2)\ln_2
  +\frac{1}{6}(\ln_2\ln_4-\ln_1\ln_3)\Bigg)
\nonumber\\&&
+m_3^2\Bigg(\frac{\pi^2}{36}+\frac{19}{54}-\frac{1}{9}\ln_3
  -\frac{m_1^2}{6\lambda_m}(\delta_m+2m_2^2)\ln_3
  +\frac{1}{6}(\ln_3\ln_4-\ln_1\ln_2)\Bigg)+\cal{O}(\epsilon)\,,
\nonumber\\&&
\ea
\ba
\label{H21p}
\lefteqn{\hskip-0.6cm
(16\pi^2)^2 \frac{\partial}{\partial p^2}H_{21}(m_1^2,m_2^2,m_3^2;0) =
-\frac{1}{24}\lambda_1
+\frac{17}{288}-\frac{m_1^2}{24\lambda_m}(\delta_m+2m_1^2)
-\frac{5m_1^4m_2^2m_3^2}{6\lambda_m^2}
}&&\nonumber\\&&
+\frac{m_1^6m_2^2m_3^2}{\lambda_m^3}
\Bigg(1+\frac{5m_2^2m_3^2}{\lambda_m}\Bigg)\Psi(m_1^2,m_2^2,m_3^2)
+\frac{m_1^6\delta_m}{12\lambda_m^2}\ln_1\Bigg(1+30\frac{m_2^2m_3^2}{\lambda_m}
  \Bigg)
\nonumber\\&&
+\frac{m_1^4m_2^2m_3^4}{\lambda_m^3}\ln_3\frac{5}{2}(m_3^2-m_1^2-m_2^2)
+\frac{m_3^2}{12\lambda_m}\ln_3   (m_3^2-m_2^2)
\nonumber\\&&
+\frac{m_1^2 m_3^2}{12\lambda_m^2}\ln_3
    \Bigg(  -3 m_1^4 -12 m_1^2 m_2^2 + 5 m_1^2 m_3^2 +
   3 m_2^4 - m_2^2 m_3^2 - 2 m_3^4 \Bigg)
\nonumber\\&&
+\frac{m_1^4m_3^2m_2^4}{\lambda_m^3}\ln_2\frac{5}{2}(m_2^2-m_1^2-m_3^2)
+\frac{m_2^2}{12\lambda_m}\ln_2   (m_2^2-m_3^2)
\nonumber\\&&
+\frac{m_1^2 m_2^2}{12\lambda_m^2}\ln_2
    \Bigg(  -3 m_1^4 -12 m_1^2 m_3^2 + 5 m_1^2 m_2^2 +
   3 m_3^4 - m_2^2 m_3^2 - 2 m_2^4 \Bigg)+\cal{O}(\epsilon)\,.
\nonumber\\&&
\ea

Here we used
\ba
\label{deflambda}
\ln_4 &=& \ln_1+\ln_2+\ln_3\,,\nonumber\\
\delta_m &=& m_1^2-m_2^2-m_3^2\,,\nonumber\\
\lambda_2 &=& \lambda_0^2+(\ln(4\pi)+1-\gamma)^2\,,\nonumber\\
\lambda_1 &=& \lambda_0+\ln(4\pi)+1-\gamma\,,\nonumber\\
\lambda_0 &=& \frac{1}{\epsilon}+\ln(4\pi)+1-\gamma\,.
\ea
These are appropriate for $\overline{MS}$ subtraction as
is customary in CHPT \cite{BCEGS1}.

Below threshold the methods of \cite{GS} lead to a two-integral
representation of the finite part
\ba
\lefteqn{
\{\overline{H},\overline{H}_1,\overline{H}_{21}\}(m_1^2,m_2^2,m_3^2; p^2)
= }   \nonumber\\
& &  \int_{(m_2+m_3)^2}^\infty d\sigma 
\sqrt{\lambda\left(1,\frac{m_2^2}{\sigma},\frac{m_3^2}{\sigma}\right)}
\times
\int_0^1 dx {\cal K}_2(x,\sigma,p^2) \{1,x,x^2\}\,,
\ea
with
\be
{\cal K}_2(x,\sigma,p^2) = \frac{1}{(16\pi^2)^2}\left(
\ln\frac{m_1^2 (1-x)+\sigma x -x(1-x) p^2}{m_1^2 (1-x)+\sigma x}
+\frac{p^2 x(1-x)}{m_1^2 (1-x)+\sigma x}\right)
\ee
and
\be
\lambda(x,y,z) = (x-y-z)^2-4 \,y\,z\,,
\ee
the K\"all\'en function. 

The dispersive representation
\be
B(m_2^2,m_3^2,t) =
\frac{\pi^{(1-d)/2}2^{3-2d}}{\Gamma\left(\frac{d-1}{2}\right)}
\int_{(m_2+m_3)^2}^\infty d\sigma
\left(\lambda\left(1,\frac{m_2^2}{\sigma},\frac{m_3^2}{\sigma}
\right)\right)^{\frac{d-3}{2}}\sigma^{\frac{d-4}{2}}
\frac{1}{\sigma-t}\,,
\ee
has been used here instead of the simpler case with equal masses
used in \cite{GS}.

Above threshold, $s\ge (m_1+m_2+m_3)^2$ the functions $\overline{H}_i$
develop imaginary parts and
they can then be evaluated from their dispersive representation
\be
\overline{H}_i(m_1^2,m_2^2,m_3^2;s) =
\frac{s^2}{\pi}
\int_{(m_1+m_2+m_3)^2}^\infty
\frac{dz}{z^2} \frac{ \mathrm{Im}\overline{H_i}(z)}{z-s}\,.
\ee
The imaginary parts are given by (in $d=4$)
\ba
\lefteqn{\mathrm{Im}
\left\{\overline{H},\overline{H}_1,\overline{H}_{21}\right\}
(m_1^2,m_2^2,m_3^2;s) =}\nonumber\\&&
\frac{-1}{16}\frac{1}{(2\pi)^3}\int_{m_1}^{E_1^{max}}d E_1
\left\{1,\frac{E_1}{\sqrt{s}},\frac{4 E_1^2-m_1^2}{3s} \right\}
\left(E_2^{max}-E_2^{min}\right)\,,
\ea
with
\ba
E_1^{max}& =& \frac{1}{2\sqrt{s}}(s+m_1^2-(m_2+m_3)^2)\,,\nonumber\\
m_{23}^2 & = & s+m_1^2-2\sqrt{s}E_1\,,\nonumber\\
E_2^{max}-E_2^{min} & = & \frac{1}{m_{23}^2\sqrt{s}}
\sqrt{\lambda(s,m_1^2,m_{23}^2)}\sqrt{\lambda(m_{23}^2,m_2^2,m_3^2)}\,.
\ea

In the text we use mainly the finite functions $H^F_i(m_1^2,m_2^2,m_3^2;q^2)$
and their derivatives with respect to 
$q^2$, $H^{F\prime}_i(m_1^2,m_2^2,m_3^2;q^2)$.
These correspond to using Eq. (\ref{expandHi}) 
and setting $\lambda_2$ and $\lambda_1$ to zero in Eqs. (\ref{H}-\ref{H21p}).

\section{Regularization and Renormalization}

In this paper we have employed the version of Modified Minimal Subtraction
(\MS)
that is customary in CHPT. The precise procedure has been discussed
in great detail in Ref. \cite{BCEGS1}.

The procedure used in Ref. \cite{KG1} corresponds to subtracting
only the $\lambda_0$ terms present in all the integrals, including those in
$\lambda_1$ and $\lambda_2$.

As mentioned in  
\cite{BCEGS1} a Taylor expansion of the $p^4$ coefficients
introduces in principle new parameters $\delta_i$ via the Laurent-expansion
of the $L_i = a_i/\epsilon + L_i^{MS} + \epsilon\delta_i+\cdots$.
We have checked that the terms involving $\delta_i$ take the form
of a local action for the quantities considered in this manuscript, 
thus they can be absorbed in the $p^6$ LECs as proven in general
in \cite{BCG}.

We have defined 
\be
\label{defLi}
L_i \equiv (\mu c)^{-2\epsilon}\left(\frac{-1}{32\pi^2\epsilon}\Gamma_i
+L_i^r(\mu)\right) =
(\mu)^{-2\epsilon}\left(\frac{-1}{32\pi^2}\Gamma_i\lambda_0
+L_i^r(\mu)+{\cal O}(\epsilon)\right)\;.
\ee 
In the main text we have suppressed the explicit $\mu$-dependence of the
$L_i^r$. The coefficients $\Gamma_i$ are given in Ref.~\cite{GL} and
$\ln c = -1/2 (\ln(4\pi)-\gamma+1)$. The order $\epsilon$ term
in the last part of Eq. (\ref{defLi}) has been used as well
to check the explicit cancellations of $\ln(4\pi)$ and $\gamma$ in all
expressions.

Similarly the coefficients in the $p^6$ Lagrangian are used to absorb
the remaining infinities via
\be
C_i \equiv (\mu c)^{-4\epsilon}\left(\frac{\gamma_{2i}}{\epsilon^2}
+\frac{\gamma_{1i}}{\epsilon}
+C_i^r(\mu)\right) =
\mu^{-4\epsilon}\left(\gamma_{2i}\lambda_2+\gamma_{1i}\lambda_1
+C_i^r(\mu)+{\cal O}(\epsilon)\right)\;.
\ee

Dropping the terms with $\lambda_0$, $\lambda_1$, $\lambda_0$,
replacing the $C_i$ by $C_i^r$ in the main text 
and subtracting the terms proportional
to $C$, $\lambda_0$, $\lambda_1$ and $\lambda_2$ in the expressions for
the integrals given in the preceding appendices,
gives the results in the \MS\ scheme.

\end{document}